# Business-cycles and Cash-on-Market: Pre-money Startup Valuation in the Macroeconomic Environment[1]

Max Berre[2] and Benjamin Le Pendeven[3]

*WORKING PAPER – PLEASE DO NOT SHARE NOR QUOTE THIS PAPER WITHOUT PERMISSION*

**Abstract**

How do business-cycles impact startup-valuations? While several studies explore VC startup-ecosystems and pre-money valuations, relatively-few delve deeper into the role of macro-level economic factors in influencing those startup deals valuations. Using a dataset of 1,089 venture-capital investments in European Union and European Economic Area markets, this article examines macroeconomic, cyclical and macro-sectoral influences on VC startups pre-money VC valuations. Our findings show that business-cycles impact startup-valuation both directly and indirectly. Beyond DCF factors, startup-valuations are impacted via by business-cycles directly, and via local venture-capital market-size. By using a Structural Equation Model approach, our findings contribute to entrepreneurship and financial-intermediary literature by exploring indirect and endogenous relationship possibilities finding that most determinants are transmission-channels rather than independent drivers. Our findings effectively tie-together startup-valuations, intermediary markets, and macroeconomic determinants.

<u>Keywords</u>: Valuation, Startup, Venture Capital, Entrepreneurial Finance, Business-cycle, Structural Equation Model

[1] This article has been sponsored by the Partners of the "Finance for innovation" Chair at Audencia Business School, especially Early Metrics and Sowefund, who provided financial support to Max Berre during his PhD course. The authors wish to express their sincere gratitude for this support. Additionally, the authors want to thank XX as editor in charge of this paper, and the X anonymous referees for their helpful comments. Moreover, thanks to Dimitris Petmezas, Aristrogenis Lazos and Paul Olivier Klein, as well as the attendees to the ENTFIN 2021, FEBS 2021, and FEM 2022 conferences for their fruitful comments.

[2] PhD Candidate, Audencia Business School, Nantes, France and Université de Lyon, iaelyon, Magellan, Lyon, France mberre@audencia.com
[3] Associate Professor, Head of "Finance for innovation" Chair, Audencia Business School, Nantes, France, blependeven@audencia.com

# 1. Introduction

In 2020 and 2021, worldwide Venture Capital (VC) markets reached records both in terms of value invested and in number of deals. In Q4 2021, VC markets hosted deal values around $191 Billion, and deal counts around 11,000 (Pitchbook, 2022). Records were also shattered in terms of startup-valuations, with for example quantity of unicorns growing from 45 in 2014 (Wall Street Journal, 2014) to 1,056 in 2021 (Hurun Research Institute, 2021). Meanwhile, experts and industry-reports document dramatic increases of valuation of funded startups, with median US pre-money valuations ranging from $11.4 Million for angel-financing to $1.22 Billion for Round-D financing and $5.08 Million for angel financing to $1.73 Billion for Round-D financing in Europe (Pitchbook, 2022). Questions of causality-structure now appear critical for practitioners (Payne, 2011; Berkus, 2016), policy-makers (Dieppe at al., 2017; Popov, 2009) and academics (Kaplan et al., 2009; Gompers et al., 2020).

Entrepreneurial finance literature related to startup valuations deeply investigated stakeholder-factors (both entrepreneur and investors sides) and deal-conditions, but dedicated limited attention to macro-environmental factors (Berre and Le Pendeven, 2022; Köhn, 2018).

Even though it has been long-understood that VC markets are deeply-procyclical investment-intermediaries, channeling investment-funds into the entrepreneurial landscape both during booms and as markets recover in aftermath of recessions, we have a limited understanding of their consequences on pre-money valuation of funded startups, after the selection stage. Heughebaert and Manigart (2012) find that more and more-successful VC-deals occur when domestic economic-growth is stronger, while Schwienbacher (2013) outlines that markets for VC financing are subject to large variations in capital-supply over the business-cycle, institutional investors channel large amounts of capital to VC-funds during booms.

As non-bank financial-intermediaries, VC-funds might be more responsive to macroeconomic conditions than is the banking-sector or many other parts of the economy. Reputational-costs on VC markets are lower during booms due to increased capital-availability (Schwienbacher, 2013), while valuation-impacts of financial variables are business-cycle dependent (Gavious and Schwartz, 2011).

In principle, this means that nuanced macroeconomic distribution-channels are at work. This phenomenon grants opportunity to draw in novel insights from wider theories by synthesizing between macro and entrepreneurial finance, a literature gap explicitly pointed-out by Budhwar et al. (2022). This gap is especially pertinent since just a couple of finance studies examined the nature of business-cycle and macroeconomic driver-impact on VC and startup markets directly. These include Fitza et al. (2004) and Korteweg and Sorensen (2010), which both empirically find business-cycle impact on valuation, a phenomenon which needs further modeling and research.

Addressing this gap is critical to understanding VC market dynamics, as well as the balance between investors and entrepreneurs' value-holdings, which gave rise to attention-grabbing headlines in recent years, as unicorns emerged in several of the world's key markets.

Using a bespoke VC-dataset consisting of 1045 European Union (EU) and European Economic Area (EEA) deals from 2000 to 2020, this study explores impacts of macroeconomic and business-cycle valuation-factors on startup-valuations. These range from macroeconomic output-gap to dry-powder, as well as country-risk-premiums and venture capital cash-on-market. Because European Union supranational decision-making and regulatory powers focus mainly on trade-policy, economic-regulation and macrofinancial-regulation (Jopp, 2017), examining EEA market-data grants rare opportunity to have insight into markets which combine divergent macroeconomic and macrofinancial landscapes with unified financial-regulatory landscapes.

We outline that many traditional valuation-drivers classically demonstrated to impact firm-valuation and startup-valuation ranging from macrofinancial indicators to intermediary-market indicators are, in fact, mediating-variables channeling business-cycle effects into the entrepreneurial landscape. Essentially, many startup-valuation drivers are actually distribution-channels rather than independent valuation factors.

Because startup-valuations are known to be driven by discounted-cashflow models (DCF), we control for discounted-cashflow-based firm and performance-characteristics in addition to market-condition valuation-factors. This leads to a viable factor-based startup-valuation model which can examine both external-information, such as sectoral market-structure or macroeconomic situation, and examinations of contextual detail, including interaction-effects and indirect-effects.

Our contributions are twofold. First, this study addresses the gap linking entrepreneurial finance and macroeconomic theory outlined by Budhwar et al. (2022). Second, our results describe the relationship between business-cycles and startup-valuations as a nonlinear relationship consisting of multiple independent valuation-channels connecting valuations with business-cycles via partially-mediating intermediate-variables. This represents a new way of conceptualizing startup pre-money valuation. Essentially, our empirical results obtained through structural-equation modeling contribute to entrepreneurship and financial-intermediary literature by describing an empirically-supported model demonstrating how business-cycles impact startup-valuations both directly and via multiple independent macro-level valuation channels.

Our paper develops as follows: section two examines literature exploring VC and startup-market impact of both country-level and macroeconomic factors, and their role in describing the determination of pre-money startup-valuations. Section three describes our dataset, as well as key-variables and empirical-approaches used to develop insight on the impact of macroeconomic and business-cycle market-conditions on startup-valuations. Section four describes empirical results of our baseline-scenario, OLS models and fixed-effects models used. Section five discusses key findings in detail, while section six outlines findings-implications, as well as conclusions and avenues for future research.

## 2. Literature Review

As demonstrated by several systematic literature reviews on pre-money startups valuations such as Köhn (2018), Wessendorf (2019), and Berre and Le Pendeven (2022), published papers in the field mostly investigate entrepreneurs, investors and deal-related factors on valuations, while macroeconomic determinants attract limited attention. In fact, the aforementioned authors describe that many papers who include macroeconomic drivers do so primarily as control variables. This means that few studies examine the role of macroeconomic market conditions in detail, thereby constituting a noticeable gap in the research. Rarer still, are studies that explore indirect effects on valuation within the entrepreneurial field.

The deterministic impact of macroeconomic and macrofinancial indicators in driving startup-investment is a long and storied idea within economic literature and can be traced to several authors, who established overall that VC markets behave as highly-procyclical financial intermediaries (Bygrave and Timmons, 1985; Korteweg and Sorensen, 2010).

**Macroeconomic Impact of Funding Availability: The VC Intermediary Role**

Venture capital markets are financial intermediary markets which are driven by business-cycles and macroeconomic conditions, even as they drive startup-valuations. As early as Chan (1983), who proposes a theoretical model outlining that investment-allocation at the macro-level depends on the VC intermediary role to manage sector-specific information-asymmetries, the intermediary role of VC markets has been well-understood. Focusing on VC fund-availability, Gompers and Lerner (1998) employ a two-stage Heckman model, estimating probability that funds are raised, followed by the amounts of funds raised. They find that while firm-characteristics have strongest effect on fundraising, commitments to VC-funds are also directly impacted by GDP-growth-rates, while capital-gains taxes affect amounts raised, but not funding-probability. Stated otherwise, the impact of VC funding is influenced by external market-conditions, while VC fund-availability is driven by growth-rates.

Meanwhile, total value of stocks traded is one of the most significant determinants in explaining divergences of VC intensity, according to Bonini and Alkan (2009), who test impacts of GDP, inflation rates, real interest rates, total number of stocks traded, as well as corporate income tax-rates and business expenditures on R&D. Other key determinants are inflation rate and total entrepreneurial activity (Bonini and Alkan, 2009), meaning that VC markets may be driven by both macroeconomic factors and also startup-market activity.

Private equity activity is driven by institutional and legal environment, as well as economic activity, unemployment rate, equity market capitalization, and unit labor costs (Bernoth and Colavecchio, 2014). By using country-level fixed-effects and dividing their sample between Western-Europe and Eastern-Europe, they find that institutional and legal environment have substantial individual explanatory-power. Overall, it is intuitive that separately-examining Eastern and Western European -markets during periods of EU-enlargement find institutional and political variables, as well as inflation to be highly-deterministic on European private equity markets. In principle, this indicates that country-level variables influence the overall impact of private equity in an economy. Nevertheless, it should be noted that while private-equity-investment measured relative to GDP indicates investor-confidence, it might not necessarily indicate returns or valuations in these markets.

Concerning funding availability, inflows of capital into VC-funds increases the valuation of their investment-targets (Gompers and Lerner, 2000), and activities in private equity markets is highly deterministic in influencing startup-valuations (Inderst and Muller, 2004; Hellmann and Thiele, 2015). In addition to cash-on-market, Inderst and Muller (2004) make several key insights, describing endogeneity between VC funds-scarcity, entry-costs, and investment-profitability, theorizing that investment-profitability drives startup-valuation as well as VC market-entry. Stated otherwise, while cash-on-market drives startup-valuations, the relationship's functional form might be subject to endogeneity and indirect valuation-impacts. Essentially, further investigation into the specific nuance of the variability of valuation-impacts gives rise to further research and debate.

**Business Cycles**

GDP-growth and business-cycle dynamics are known to play roles in determining startup-valuation (Dieppe at al., 2017). For example, more and more-successful VC deals occur when Belgian economic growth is stronger (Heughebaert and Manigart, 2012). On a general level, Dieppe at al. (2017), a European Central Bank (ECB) study describing the ECB's global macroeconomic spillover-analysis model, describe financial-variables as key-determinants of domestic-output, with financial-spillover-transmission occurring via four channels: equity-prices, interbank-rate-spread, lending-tightness, and sovereign-risk-premium relationship. Lending-tightness in turn is driven by momentum and expected-GDP-growth. The intermediary-role played by lending-tightness, having deterministic-effect on the real-economy, while itself being impacted by key macroeconomic variables is informative to understanding structural relationships between macro-level variables and the real-economy.

Dées (2016), another ECB study uses a GVAR model-approach to describe transmission of financial-shocks to the real-economy, finds that financial-variable impact on the real-economy remains broadly-unchanged over time, as the business-cycle progresses. This finding of financial-variable impact-stability is informative to understanding structural relationships surrounding the impact on valuations.

Nevertheless, VC-financing markets are subject to large variations in capital-supply over the business-cycle, seeing large inflows from institutional-investors during boom-periods (Schwienbacher, 2013), while financial-integration drives business-cycle-synchronization via indirect-effects, by promoting trade-integration and economic structural-similarity, even as direct qualitative and quantitative financial-integration-effects indicate otherwise (Trancoso and Gomes, 2020). This study is highly-relevant due to its separation of direct and indirect impacts on business-cycles, demonstrating the inherent complexity of the relationship between business-cycles and the overall economic situation, demonstrating the possible need for a structural-equation-model approach.

Literature describing direct and indirect impacts on business-cycles, according to potentially-complex causal-architecture gives rise to the question of what exactly this architecture actually looks like. The relationship between direct and indirect determinants in this area is one seldom investigated.

**Macrofinancial Interaction Effects**

Not only do macroeconomic and business-cycle indicators play a key deterministic role in startup markets, but also that interaction-effects between macroeconomic indicators and industry-level indicators also likely play a role, demonstrating that VC markets play a highly-variable financial intermediary role.

Popov (2009) corroborates this by focusing on the relationship between VC markets and firm-size, proposing a model which describes firm size as a product of industry-level firm-sensitivity to VC finance and to bank finance, as well as to VC market-size. This author highlights that industry-level sensitivity to VC markets drives firm-level employee size and establishes that relationships exist not only between financial-market-conditions and startups, but also that this relationship is industry-specific (ie, that this relationship is influenced by sectoral variables and is subject to multivariate functional-form).

Meanwhile, firm characteristics such as EBITDA are weighed differently from country to country (Lockett et al., 2002), while lower risk-free-rate drives overvaluation, while illiquidity premium reduces overvaluation, meaning fundamentally that valuation discount-factors are driven by overall cyclical and macrofinancial conditions (Gornall and Strebulaev, 2020). This demonstrates that the valuation-effect of DCF valuation-factors (i.e., revenues and discount-factors) are impacted by macroeconomic and cyclical market-conditions. This however raises question of specifically where and how these effects occur.

While the existing literature establishes the relationship between VC markets, inputs to startup-value (ie, firm and entrepreneur-characteristics) and startup valuations, is influenced by market-conditions via direct, indirect, and interactive influence from sectoral and macroeconomic conditions, whose functional form is complex and bears further investigation.

**Background and Research Hypotheses**

Market-conditions are macro-level indicators, to sectoral and cyclical indicators. Hsu (2007), who uses year-dummies, observes that valuations tend to be higher when fund inflows are high, as well as Armstrong et al. (2006), who control for a benchmark financial-index, finding its valuation-impact positive and significant. In addition, interaction-effects between market-conditions and firm-characteristics can create additional valuation-synergies. Examples in literature include Gavious and Schwartz (2011), who find that the valuation-impact of a firm's balance sheet financials evolves along with the business-cycle, growing in importance during a post-bubble recession. In a similar vein, Gompers et al. (2008) find that interaction effects exist between sectoral market conditions and investor characteristics.

At a general level, startup-valuation may be driven by overall economic and cyclical activity. Several studies suggest this to be the case: for example, more and more-successful VC deals occur during boom periods (Heughebaert and Manigart, 2012). Moreover, business-cycle impacts on valuation (Fitza et al., 2004; Korteweg and Sorensen, 2010), while overall economic activity directly impacts private-equity activity's share of the macroeconomy (Bernoth and Colavecchio, 2014).

While there are several ways that business-cycles can directly impact valuation, both revenues and discount-factors (i.e., beta, country-risk-premium) are controlled for directly. Business-cycles might directly impact valuations due to potential future revenue-growth, industry-growth, or macroeconomic-growth, all of which may serve to attract investors, influencing both investor selection and startup-valuation (Wessendorf, 2019). Meanwhile, business-cycle growth can impact valuations by means of growth of business networks, partnerships, and relationships (Streletzki and Schulte, 2013). Alternatively, macroeconomic business-cycle conditions can also serve to inflate startup-valuations in the short-run during a boom, albeit on a short-lived based which may later be corrected, as described by Michel (2014). Thus, it can be expected that macroeconomic indicators drive not only overall activity but also valuations.

On the other hand, the valuation of startups can be driven by market-conditions more-specific to VC markets. Studies elaborate that not only does venture capital cash-on-market directly a priori influences pre-money valuations, as described by Inderst and Muller (2004), but also influences entrepreneur-effort, as described by Fulghieri and Sevilir (2009), which both increase valuations directly, and also serve as value-signal to investors. For European-market deals captured by our dataset, this can be tested by measuring the impact of business-cycles, macroeconomic indicators, and country-level VC-sector size on startup-valuations. This leads us to:

Hypothesis 1: Market Conditions. Macro-level conditions or financial-intermediary conditions

> *H1a: The macro drives the market. Startup-valuations are driven by macroeconomic, macrofinancial, and macro-level indicators. Because are the basis for revenue and discount-rate projections, direct valuation-impacts exist.*

> *H1b: Startup-valuations are driven by VC cash-on-market. Domestic-cash-on-market impacts indicate relatively-insular VC markets. Global-cash-on-market impacts indicate cross-border valuation-effects.*

Meanwhile, these conditions on VC markets may themselves be driven by the same business-cycle conditions impacting startup-valuations. Several studies describe that non-bank financial intermediary markets (which includes both VC markets and other investment markets) are influenced by business-cycles in a procyclical manner, with macroeconomic activity impacts overall private-equity activity, meaning that as a macroeconomy grows, investment in its emergent firms grows as well (Bernoth and Colavecchio, 2014).

Financial-spillover-transmission occurs via equity-prices, interbank-rate-spread, lending-tightness, and sovereign-risk-premium relationships. Fundamentally, this means that cash-on-market likely transmits financial spillovers of the business-cycle originating elsewhere in the wider macroeconomy (Dieppe at al., 2017). Meanwhile, lending is subject to macroeconomic-level variations driven by (macro-level) tail-risks (Chu and Zhang, 2022), while VC-financing markets are subject to substantial variations in capital-supply over the business-cycle (Schwienbacher, 2013). In conjunction with Dieppe at al. (2017), this means that VC markets function as a manifestation of financial-spillover-transmission channel, in line with equity-prices and lending-tightness. Because cash-on-market captures VC-sector size in a country, it bears testing whether it is deterministically-influenced by macroeconomic and business-cycle indicators, leading us to:

Hypothesis 2: The Financial Intermediaries Market. Cash-on-Market

> *H2: Cash-on-market is driven by macroeconomic and cyclical-indicators. Because direct valuation-impact of both cash-on-market and macrofinancial indicators exist, this may create a second-order valuation-impact effect.*

In principle, while startup-valuations are known to be driven by revenue and by parseable risk-premiums (Damodaran, 2009), the *manner in which* and *degree to which* they impact startup-valuations can themselves be country or industry-specific (Damodaran, 2010), or else influenced by conditions in countries, cities, and industries. Specifically, the nature and scale of the impact of DCF valuation-factors can be influenced by macro-level market conditions, which might be either regulatory or macroeconomic in nature (Gompers and Lerner, 1998; Lockett et al., 2002; Bernoth and Colavecchio, 2014). Alternatively, they can be influenced by industry-level effects (Chan, 1983; Damodaran, 2009), Corroborating this, Manigart et al. (1997, 2000), identify that valuation-impact-differences between sectors, geographic settings, and investor-contexts. On the other hand, alternate views describe that national-level economic-effects can be driven by local-level, investor-side, or supply-chain effects (Porter, 1990). Combinations of these factors impacting the relationship between startup-valuation and valuation-drivers have also been observed for VCs across different countries focusing on fintech-industry startups.

In terms of functional-form, what this describes are divergences in the impact of DCF valuation factors across cities, countries, industries, and investors, or joint-combination thereof, which might be specific to those industries, cities, investors, and countries, or whose valuation-divergences might be driven by these differences. Functionally, industry fixed-effects can explain differences in valuations by capturing industry level business-relationships, business models and firm-survivability differences (Damodaran, 2009).

Meanwhile, country-level fixed effects may capture national-level differences not only in macroeconomic fundamentals, but also in governance and economic policy differences (Gompers and Lerner, 1998; Bernoth and Colavecchio, 2014). Governance and policy differences lead to divergences in valuation approaches and in relative-importance of valuation-drivers (Rojo-Ramirez, 2013). Lastly, city fixed-effects can capture economic clustering, as well as city-level investor and industry-concentration (Porter, 1990).

Fundamentally, since literature describes that impact of DCF valuation-factors can be influenced by country-level, industry-level or local-level dynamics, it is important to not only compare these, but also to examine the specific nature of the functional form by which each of these has their specific influence on the valuation-impact of the DCF valuation-factors. In principle, this can be measured by controlling for categoricals such as country, industry, and city, via fixed-effects, leading us to:

Hypothesis 3: Subsets Among Startups and VCs

> *H3: The macroeconomic situation affects VC investors and startups in some subsets of the dataset more than others. Therefore, fixed-effects regressions may tell a dramatically different story than baseline macroeconomic and macro-level regressions.*

Concerning the influence of market-conditions on the valuation-impact of DCF-factors, an alternate approach to the functional-form question models indirect effects and relationships. The role of business-cycles – in particular booms – on the valuation impact of other variables, ranging from financial variables (as described by Gavious and Schwartz, 2011) to reputational costs (as described by Schwienbacher, 2013), indicate clearly that business-cycle has valuation-impacts have not only direct, but indirect valuation-impact as well, with business-cycle influencing intermediate valuation-factors such as VC-industry variables, country-risk-premiums, and macroeconomic determinants to further influence valuation.

Our study is far from the first to model indirect effects. Empirical literature outlines that formal entrepreneurship is driven by availability of economic opportunities, which is in-turn driven by macro-level resource-availability (Thai and Turkina, 2014), while financial-variables act as key-determinants of domestic-output, with financial-spillover-transmission occurring via four channels, equity-prices, interbank-rate-spread, lending-tightness, and sovereign-risk-premium relationship, with lending-tightness in turn being driven by momentum and expected-GDP-growth (Dieppe at al., 2017). In principle, the overall relationship between startups, macroeconomic-cycles, and intermediary financial valuation-impact channels likely follows a similar pattern.

This pattern is explored by Dées (2016), using a GVAR model-approach to capture the relationship between valuations and intermediate variables, which might indicate endogenous fully or partially-circular causal-relationships. A similar pattern is corroborated by Trancoso and Gomes (2020), who examine direct and indirect-effects linking financial-indicators and business-cycle synchronization, affirms that structural-equation-model approaches are particularly well-suited to examinations of business-cycle dynamics and relationships. Additionally, beyond capturing indirect effects, SEM-approaches facilitate capture of endogenous fully or partially-circular relationships among dependent, independent, and mediating-variables.

Essentially, a group of studies lead us conclude that indirect-effects between valuations, intermediate-variables and wider macro-cyclical conditions can be captured by multistage econometric models (Thai and Turkina, 2014; Dées, 2016; Trancoso and Gomes, 2020). What remains to be seen is the precise shape and architecture of the direct and indirect relationship, as well as whether any parts of the relationship-architecture are endogenous in nature. In principle, relationship-architecture shape can be tested via structural-equation model techniques, so that the specific structure of indirect-effects can be modelled in detail, leading us to:

Hypothesis 4: Direct and Indirect Effects. Structure and Functional Form

> *H4a: Business-cycles and financial-intermediary market-conditions impact startup-valuations directly and independently of one-another.*
>
> *H4b: Business-cycle-impacts on startup-valuations are non-circular full or partial-mediation effects, impacting valuations both directly (in the case of partial-mediation), and via other market conditions.*
>
> *H4c: Business-cycle-impacts on startup-valuations are circular full or partial-mediation effects, impacting startup-valuations with an endogenous circular relationship between mediating financial-market conditions and startup-valuations.*

## 3. Data and Methodology

In this section, we outline key modelling approaches that are used to capture and describe indirect and contextual-effects the literature describes market conditions as having on valuation-impact of the DCF-related valuation-drivers. While literature describes indirect and contextual impact on the effect of these valuation-determinants, the specific nature of the relationship has remained a matter of debate.

**Scorecard-Based Valuation**

Pre-money startup-valuations are prices (or prices of their shares) paid by equity investors for early-stage investments (Cumming and Dai, 2010). Scorecard-valuation methods are modular and relatively straightforward valuation-approaches based on summation of key characteristics, market-conditions, and deal-conditions developed mainly by industry practitioners. Meanwhile, in the economic literature, this same concept appears as summation-based valuation models (Hand, 2005; Sievers et al., 2013). In principle, scorecard approaches can model indirect valuation-influences in several ways, as well as incorporating non-financial and deal-characteristics prevalent in a given market or subset thereof.

**Startup-Valuation Meta-Model**

Berre and Le Pendeven (2022) develop a model outlining the contextually-adjusted startup-valuation-process, assigning factors to appropriate process-chronological-positions. Essentially, startup-value-inputs navigate external market-conditions as part of the valuation-forming process. This implies both direct-effects and interaction-effects of market-conditions on startup-valuation.

*Equation 1: Berre-Le Pendeven Startup-Valuation Meta-Model*

$$Pre-Money\ Valuation = f(((\sum Startup\ Value) \sum Deal\ Value) \sum Deal\ Valuation)$$

In order to approach a clear view on the explanatory power and valuation-impact of macroeconomic factors and macro-financial factors in the startup ecosystem, factors used in classical firm valuation models – in particular, DCF-related factors –should be included empirical models. The inclusion of classical factors serves two purposes. First, their inclusion can be used to test and establish the conceptual and theoretical soundness of our dataset. Second, inclusion of classical valuation-factors serves as key control factor that need to be taken into account, in order to isolate macroeconomic and macro-financial effects.

Subsequently, our empirical approach focuses on the examination of the startup valuation-impact of key macroeconomic and macrofinancial factors, including both indicators and policy factors. These include macroeconomic output-gap, macro-level tax-rates, and total cash-on-market levels available both domestically and on the world VC market.

## Structural-Equation-Model-Approach

Scorecard-valuation functional form notwithstanding, accurate-measurement of valuation-impacts of market-conditions face two principal modeling-obstacles: endogeneity and indirect-effects (e.g. partial and full-mediation-effects). In order to address these obstacles, we employ partial-least-squares structural-equation modeling (SEM), an approach which has several relevant advantages given the context examined. Adopting an SEM approach allows us to analyze complex indirect-relationships (Gefen et al., 2000; 2011), which feature both direct and indirect-effects (Little et al., 2007), and which are appropriate for exploratory theory-building and prediction. As a nonparametric estimation-technique (Wold, 1982), this approach provides an iterative combination of empirical-analyses that relates measures to constructs and a path analysis that captures the structural model represents the direct and indirect relationships among key valuation-drivers.

## Indirect-Effects

Business-cycle impact on startup-valuation is as likely to be indirect, acting via effects on financing-costs and conditions, as it is direct, acting directly on firm-revenues, business-opportunities, local-market-demand, and firm-level asset-values, all of which tie the firm's fortunes directly to the business-cycle.

Structural-equation-model-approaches allow simultaneous measurement of direct and indirect-effects, including full, partial, and inconsistent mediation, as well as latent and unobserved variables, while observed relationships might be part of a complex, qualified relationship-system (Little et al., 2007). Structural-equation-model-approaches are particularly well-suited to investigating business-cycle-effects, dynamics, and drivers (Trancoso and Gomes, 2020).

## Endogeneity

Wooldridge (2010) outlines that in principle, endogeneity arises due to one of three causes; omitted-variable-bias, simultaneity, or measurement-error. In principle, endogeneity-testing can indicate whether the causal-relationship is either more-complex than detected at-first-glance, or whether the causal relationship is influenced by either additional explanatory-variables or measurement-error.

While directly-circular endogenous relationships between business-cycle and startup-valuation are unlikely, given a dataset dominated by startups in large, historically-established European economies, relationships between business-cycles, cash-on-market, and startup-valuation may give rise to circular cash-on-market-relationships whereby business-cycles and cash-on-market may be endogenously-related. Additionally, cash-on-market might be endogenously-related to valuation.

Alternately, endogeneity might be driven by hidden-variable-bias (Wooldridge, 2010), which might be found acting upon the model's dependent-variables, its explanatory-variables, or both. While use of two-stage-least-squares is a well-established model-approach in accounting for endogeneity, selection of the variable, as well as relationships with both dependent-variables and other explanatory-variables must be given careful consideration.

Addressing these concerns, structural-equation models are proposed in Figure 1. In principle, two-Stage least squares regression analysis, used in the analysis of structural equations (Gefen et al., 2011), is essentially an extension of the OLS method, useful when the dependent-variable's error terms are correlated with the independent-variables (Wooldridge, 2010).

In Figure 1, valuation-impact of direct business-cycle effects, as well as indirect-effects via financing-conditions are modeled. Panel A describes the business-cycle-valuation relationship as a non-circular relationship subject to partial-mediation via financing conditions, while Panel B describes the business-cycle-valuation relationship subject to partial-mediation via financing conditions and with endogenous relationships between business-cycle and cash-on-market and between valuation and cash-on-market.

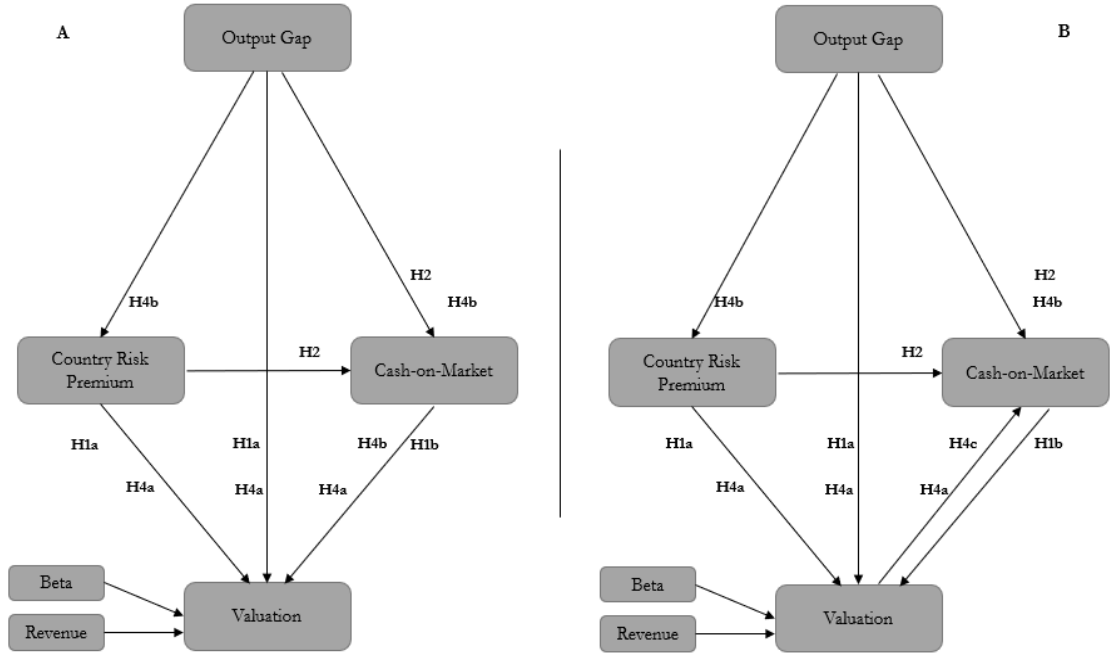

*Figure 1: Structural-Equation-Model Valuation-Relationship Path-Diagram: Non-Circular and Circular Models*

**Equation Structure**

In terms of equation-structure, Figure 1A describes a two-tiered structural-equation model subject to partial-mediation involving both country-risk-premium and cash-on-market as mediating-variables, as follows:

*Equation 2: Output-Gap and Startup-Valuation*

$$[1]\ Cash-on-Market_t = \sum_{t=1}^{T}\beta_{1t}(Output\ Gap_t) + \sum_{t=1}^{T}\beta_{2t}(Credit-Risk-Premium_t) + \varepsilon_{1t}$$

$$[2]\ Valuation_t = \sum_{t=1}^{T}\beta_{3t}(Revenue_t) + \sum_{t=1}^{T}\beta_{4t}(Sectoral-Risk-Beta_t) + \sum_{t=1}^{T}\beta_{5t}(Credit-Risk-Premium_t) + \varepsilon_{2t}$$

While circular-relationships between business-cycle indicators and cash-on-market might be possible in principle, they are both unlikely given the composition of our dataset (i.e., deals in established and highly-industrial EEA markets, whose VC markets are comparatively smaller as a share of GDP, and whose business-cycles are often driven by European-level cyclical transmission channels (Weyerstrass et al., 2006), and difficult-to-measure, given the need for an instrumental-variable which would impact business-cycles but not cash-on-market. On the other hand, partial-endogeneity, involving an endogenous relationship between Valuation and Cash-on-Market, is both measurable and supported by economic theory. In terms of equation-structure, Figure 1B describes a partially-circular relationship, featuring an endogenous relationship between cash-on-market and contemporaneous-endogeneity of startup-valuation, as follows:

*Equation 3: Output-Gap and Startup-Valuation under partial endogeneity*

$$[1]\ Cash-on-Market_t = \sum_{t=1}^{T} \beta_{1t}(Output\ Gap_t) + \sum_{t=1}^{T} \beta_{2t}(Valuation_t) + \sum_{t=1}^{T} \beta_{3t}(Credit-Risk-Premium_t) + \varepsilon_{1t}$$

$$[2]\ Valuation_t = \sum_{t=1}^{T} \beta_{4t}(Revenue_t) + \sum_{t=1}^{T} \beta_{5t}(Sectoral-Risk-Beta_t) + \sum_{t=1}^{T} \beta_{6t}(Credit-Risk-Premium_t) + \varepsilon_{2t}$$

**Dataset**

The first piece of the dataset consists of proprietary VC deal-data shared by Early Metrics, a Paris-based startup ratings and research agency (80 observations). To this, we add EU-and EEA-located startup-deals drawn from EIKON and Crunchbase (for the valuation disclosed deals, respectively 397 and 614 observations). These supplementary sources were chosen due to their content-similarity to Early Metrics data and regular use in entrepreneurial finance researches. To further enrich the dataset, each deal was cross-referenced with firm-performance, industry-level, municipal, and national-level macroeconomic data, and including both proprietary, and commercially-available data, while also boasting extensive variety of value-adding categorical-variables. While our categorical-variables have substantial explanatory-power in their own right, they also add value by virtue of their interaction with both firm-characteristics, as well as macroeconomic and business-cycle market-conditions that are this study's focus.

Selection of European data gives our study numerous strengths, ranging from institutional and macroeconomic diversity sufficient for meaningful geographic fixed-effects to market-examination taking into taking into account of distinct contextual and geographic factors.

Because each line within our dataset is specific per-investor-per-deal, deals with multiple investors occupy multiple lines within the dataset, identifying data for startup and investor. Since a startup can have several investors, it can have multiple observations in the regression analysis, reflecting each unique investor–startup pair. This structure is borrowed from Masulis and Nahata (2009).

With 1,089 observations representing 1,042 deals across 673 startups ranging from Q1-2000 to Q1-2020, our dataset-size is substantial, although only 582 observations contain firm-level revenue figures. Nevertheless, this yields regressions with substantial degrees of freedom compared to prominent studies in the entrepreneurial finance and startup field, such as Masulis and Nahata (2009), Greenberg (2013), and Gompers et al. (2020), who examine 273, 317, and 444 observations respectively.

*Table 1a: Dataset*

| Source | Observations |
|---|---|
| EIKON | 397 |
| Early Metrics | 78 |
| Crunchbase | 614 |
| **Total** | **1089** |

**Dependent Variable**

The primary dependent variable used in this study is pre-money startup-valuation. This reflects the product of share price before a funding round multiplied by the number of outstanding startup shares. Since the dataset is drawn from EU and EEA data, valuations are expressed in EUR. Data drawn from outside the Eurozone, such as from the UK, Poland, Norway, Sweden, and Switzerland were converted into EUR. Table 1b outlines the summary statistics of our pre-money valuations data. While the data's time and sectoral distribution is somewhat uneven, it does cover several major events, including the end of the dotcom bubble, the Eurozone crisis, and the start of the Covid-19 Pandemic.

*Table 2b: Summary Statistics*

| | 2000 | 2001 | 2002 | 2003 | 2004 | 2005 | 2006 | 2007 | 2008 | 2009 | 2010 | 2011 | 2012 | 2013 | 2014 | 2015 | 2016 | 2017 | 2018 | 2019 | 2020 | Totals |
|---|---|---|---|---|---|---|---|---|---|---|---|---|---|---|---|---|---|---|---|---|---|---|
| Mean Valuation | 109 000 000 | 49 700 000 | 32 900 000 | 21 800 000 | 50 800 000 | 12 000 000 | 25 700 000 | 15 200 000 | 4 181 531 | 511 000 000 | 11 900 000 | 33 300 000 | 40 100 000 | 10 300 000 | 65 400 000 | 182 000 000 | 90 100 000 | 298 000 000 | 286 000 000 | 1 130 000 000 | 1 170 000 000 | 222 000 000 |
| Std. Dev | 239 000 000 | 155 000 000 | 29 800 000 | 15 500 000 | 29 400 000 | 7 679 114 | 32 100 000 | 9 378 568 | 3 352 995 | 690 000 000 | . | 9 424 469 | 135 000 000 | 36 100 000 | 234 000 000 | 363 000 000 | 232 000 000 | 634 000 000 | 787 000 000 | 1 300 000 000 | 1 420 000 000 | 602 000 000 |
| Min | 795 216 | 83 579 | 57 287 | 6 969 987 | 30 000 000 | 762 500 | 2 970 006 | 1 708 426 | 762 500 | 1 558 441 | 11 900 000 | 9 999 972 | 55 714 | 53 833 | 50 000 | 63 330 | 63 330 | 80 000 | 2 160 000 | 1 412 881 | 2 354 802 | 50 000 |
| Max | 1 490 000 000 | 1 280 000 000 | 120 000 000 | 43 700 000 | 71 500 000 | 16 500 000 | 48 400 000 | 24 000 000 | 8 899 903 | 1 270 000 000 | 11 900 000 | 36 700 000 | 556 000 000 | 229 000 000 | 1 560 000 000 | 1 920 000 000 | 1 920 000 000 | 2 140 000 000 | 2 230 000 000 | 5 380 000 000 | 5 060 000 000 | 5 380 000 000 |
| Austria | - | 3 | - | - | - | - | - | - | 2 | - | - | - | - | - | 1 | - | - | - | - | - | - | 6 |
| Belgium | - | 3 | - | - | - | - | - | - | - | - | - | - | - | - | - | 1 | - | - | 3 | 5 | 5 | 17 |
| Croatia | - | - | - | - | - | - | - | - | - | - | - | - | - | - | - | - | - | - | - | - | - | 1 |
| Czech | 6 | - | - | - | - | - | - | - | - | - | - | - | - | - | - | - | - | - | - | - | - | 6 |
| Denmark | - | 3 | - | - | - | - | - | - | - | - | - | - | - | 1 | - | - | - | 1 | - | 3 | - | 8 |
| Finland | 1 | - | - | - | - | - | - | - | - | - | - | 7 | - | 1 | 1 | 5 | 2 | 4 | 1 | - | - | 22 |
| France | 29 | 11 | 4 | - | - | 5 | - | - | - | 2 | - | 1 | 1 | 4 | - | 5 | 27 | 16 | - | 1 | - | 106 |
| Germany | 19 | 11 | 3 | - | 1 | - | - | - | 2 | - | - | - | 2 | 3 | 5 | 12 | 14 | 33 | - | 31 | 2 | 138 |
| Ireland | - | 1 | - | - | - | - | - | - | - | - | - | - | - | - | 1 | 5 | 8 | 5 | - | - | - | 20 |
| Italy | 3 | - | - | - | 1 | - | - | - | - | - | - | - | 1 | 1 | - | 1 | 2 | 1 | - | - | - | 10 |
| Latvia | - | - | - | - | - | - | - | - | - | - | - | - | - | - | - | 1 | - | - | - | - | - | 1 |
| Lithuania | - | - | - | - | - | - | - | - | - | - | - | - | - | - | - | - | - | - | - | 7 | - | 7 |
| Luxembourg | 1 | - | - | - | - | - | - | - | - | - | - | - | - | - | - | 2 | 3 | - | - | - | - | 6 |
| Netherlands | 2 | 1 | 2 | - | - | - | - | - | - | - | - | - | - | - | - | 5 | - | - | 1 | - | - | 11 |
| Norway | 2 | 4 | - | - | - | - | - | 2 | - | - | 1 | - | - | - | - | - | - | - | 1 | - | - | 10 |
| Poland | 9 | 8 | - | - | - | - | - | - | - | - | 1 | - | - | - | 1 | 1 | 2 | - | - | - | - | 22 |
| Portugal | - | - | - | - | - | - | 1 | 1 | - | - | - | - | - | - | - | 1 | 2 | 2 | 1 | - | - | 8 |
| Romania | 4 | - | - | - | - | - | - | - | - | - | - | - | - | - | - | - | - | - | 2 | - | - | 6 |
| Spain | 5 | - | 1 | - | - | - | - | - | - | - | - | - | - | 2 | 1 | 2 | 4 | 2 | - | 4 | - | 21 |
| Sweden | 5 | 3 | - | - | - | - | - | - | - | - | - | 1 | 1 | - | 1 | 13 | - | 2 | - | 1 | - | 27 |
| Switzerland | 4 | 1 | 1 | - | - | - | - | - | - | - | 1 | - | - | - | 1 | - | 2 | 3 | - | - | - | 13 |
| UK | 43 | 59 | 6 | 4 | - | 2 | 1 | 2 | 1 | 1 | - | - | 12 | 30 | 89 | 151 | 116 | 54 | 5 | 26 | 21 | 623 |
| Total | 133 | 108 | 17 | 4 | 2 | 7 | 2 | 5 | 5 | 5 | 1 | 9 | 17 | 42 | 106 | 200 | 182 | 130 | 12 | 73 | 28 | 1089 |
| Business / Consumer Services | - | - | - | - | - | - | - | 1 | - | - | - | - | 5 | 8 | 30 | 33 | 21 | 12 | - | - | - | 110 |
| Aerospace | - | - | - | - | - | - | - | - | - | - | - | - | - | - | - | - | - | 1 | - | - | - | 1 |
| Retail | 12 | 9 | - | - | - | - | - | - | - | - | - | - | 1 | 3 | 9 | 22 | 31 | 2 | - | 7 | - | 96 |
| Automotive | - | 2 | - | - | - | - | - | - | - | - | - | - | 1 | 1 | - | 2 | 9 | - | - | - | - | 15 |
| Finance | 5 | 4 | 2 | - | 1 | - | - | 1 | - | 1 | - | - | 1 | 4 | 5 | 45 | 23 | 23 | 1 | 17 | 4 | 137 |
| Food/Agro | 3 | - | - | - | - | - | - | - | - | - | - | - | - | - | 4 | 16 | 12 | 16 | 16 | 1 | - | 69 |
| Machinery / Industrial | 6 | 4 | 1 | 2 | - | - | - | - | - | - | - | - | - | 1 | - | 2 | 5 | 3 | - | - | - | 24 |
| Power | 1 | 1 | - | - | - | - | - | - | - | - | - | - | 4 | - | - | - | - | - | - | - | - | 6 |
| ICT / Software | 62 | 45 | 6 | 1 | - | 5 | 2 | 2 | 3 | - | - | 9 | 5 | 9 | 23 | 27 | 35 | 33 | 5 | 36 | 18 | 326 |
| Pharma / Healthcare | 18 | 21 | 7 | - | 1 | 2 | - | - | 2 | 1 | 1 | - | - | 1 | - | 11 | 13 | 3 | 1 | 1 | 5 | 88 |
| Education | - | 1 | - | - | - | - | - | - | - | - | - | - | 1 | - | 1 | 3 | 4 | 3 | 1 | - | - | 14 |
| Electronics | 2 | 1 | - | - | - | - | - | - | - | - | - | - | 1 | 2 | - | 3 | 2 | 1 | 1 | - | - | 13 |
| Leisure / Entertain/ Tourism | 5 | 3 | 1 | - | - | - | - | - | - | - | - | - | 2 | 2 | 10 | 15 | 18 | 5 | - | 3 | - | 64 |
| Clean-Tech | - | 1 | - | - | - | - | - | 1 | - | - | - | - | 1 | - | - | 1 | 1 | 4 | 2 | - | - | 11 |
| Home | - | - | - | - | - | - | - | 1 | - | - | - | - | - | - | - | 1 | - | - | - | - | - | 2 |
| Real Estate | 1 | 2 | - | - | - | - | - | - | - | - | - | - | - | - | 1 | 11 | 3 | 8 | - | 9 | - | 35 |
| Office | - | - | - | - | - | - | - | - | - | - | - | - | - | - | - | 2 | - | 1 | - | - | - | 3 |
| Fossil | - | 1 | - | 1 | - | - | - | - | - | - | - | - | - | - | - | - | - | - | - | - | - | 2 |
| Transport | 1 | - | - | - | - | - | - | - | - | 2 | - | - | - | 2 | 1 | 11 | 6 | 6 | - | - | - | 29 |
| Media | 7 | 3 | - | - | - | - | - | - | - | - | - | - | - | 1 | 8 | 1 | - | - | - | - | - | 20 |
| Total | 123 | 98 | 17 | 4 | 2 | 7 | 2 | 5 | 5 | 5 | 1 | 9 | 17 | 42 | 105 | 200 | 180 | 130 | 12 | 73 | 28 | |

**Independent Variables: Macroeconomic and Macrofinancial Market-Conditions**

In line with Gompers and Lerner (1998) and Bonini and Alkan (2006), which find valuations to be driven by macroeconomic indicators, cyclical indicators and tax-rates, this study's macroeconomic data also consist of country-risk-premiums, cyclical indicators and tax-rates. While GDP growth would hypothetically constitute a viable way to keep track of business-cycle conditions, our dataset, which consists mainly of recessionary-period data (which features positive, but below-trend GDP-growth figures, as well as having smaller deterministic effect during booms than during recessions), needs cyclical-contextualization in order to be a reliable explanatory variable. Therefore, we select macroeconomic output-gap, which is drawn from OECD figurers as our primary macroeconomic market-condition indicator. Additionally, at the macroeconomic level, we include an IMF tax-rate figure, tax revenue as a share of GDP, as well as country-risk-premium, drawn from the NYU-Stern database. Because country-risk-premiums are core-components of risk-adjusted discount-rates (Damodaran, 2009), they are also included in the baseline-scenario, which is driven by discounted-cashflow valuation models.

To this, we add total VC market-size available at both country level and on global markets, which Inderst and Muller (2004) describe as being a key determinant of startup valuations, as well as dry powder, both of which may act as intermediate variables in the valuation process. The former are drawn from OECD annual data on total VC investments per country, Meanwhile, the latter are drawn from Invest Europe, which provides annual dry-powder figures for six multi-country regions in Europe. Subsequently, national-level annual dry-power figures are estimated on the basis on each country's fraction of total VC investments in that region.

As an alternative to capturing business-cycle with macroeconomic output-gap, we also use Eurozone Growth Cycle Coincident Indicator (GCCI) as a robustness-check, whose results are outlined in Appendix III.

**Valuation-Driver Interaction**

In addition to direct valuation-impacts of DCF-related valuation-factors, as well as macro-level drivers on startup-valuation, impacts those variables have on one-another demand consideration.

The overall impact of economic-indicators can vary in as business-cycles progress (Blanchard and Leigh, 2013). Notable examples of these sorts of impacts existing within published entrepreneurial finance literature include Gompers et al. (2010), which find interaction effects between investor-experience and macroeconomic indicators, as well as Gavious and Schwarz (2011), who find evidence of interaction-effects between business-cycle and a startup's firm characteristics.

This means regressions including interaction-effects are justified by both theory and existing literature. To this end, interaction variables have been established for the potential interaction of each of the DCF-related variables with cash-on-market, and business-cycle.

**Baseline Model: Control Scenario**

Classically, depending on the valuation-approach, valuations are driven by firm-performance and revenue indicators, such as sales revenue, net income, cash flow from operations, EBIT or EBITDA, discounted by risk-factors such as industry-risk-premiums, country-risk-premiums, WACC, and CAPM-betas, as per DCF models, contextualized by balance sheet or income statement figures, as per multiples valuation models, and moderated by growth-rates such as in the Gordon growth model (Damodaran 2005; 2009; 2010).

Functionally, the DCF-approach can be approximated using OLS-regressions by regressing valuation against revenue indicators, as well as DCF discount-factors, such as sector-level unlevered-beta, and country-risk-premium. Fundamentally, because Damodaran holds that existing valuation approaches either apply DCF directly, or else borrow from DCF assessments caried out by others (Damodaran 2005; 2009; 2010), as with relative and multiples valuation approaches, valuation-impact of DCF valuation-determinants need to be controlled-for, both as a means of establishing the theoretical-consistency of the dataset, as well as to determine the concrete value-added of non-DCF valuation factors such as macroeconomic market conditions, financial-intermediaries market conditions, and business-cycle conditions.

While startup revenues are included in the Early Metrics deals dataset, as well as a minority of deals drawn from EIKON, additional revenue figures were drawn Dun and Bradstreet as well as Zoominfo, thereby filling-in revenue figures for approximately half of the dataset.

Sectoral unlevered-beta, an industry-level measure of systemic-risk, is drawn from Damodaran's NYU-Stern dataset, and communicates industry-level sensitivity to financial-market benchmark-volatility. These figures are used by Damodaran (2009), as inputs for risk-adjusted discount-rates used in DCF-based valuation-models.

*Country-risk-Premium* is drawn from Damodaran's NYU-Stern dataset's Country Default-Spreads and Risk-Premiums page, which was accessed on 31 January 2021. While this, in principle represents sovereign bond ratings and appropriate default spreads for different countries, these figures are used by Damodaran (2009), as inputs for risk-adjusted discount-rates used in DCF-based valuation-models.

While DCF-related valuation-factors are used to construct DCF-based regressions to test dataset-soundness, for subsequent regression-analysis, DCF-related regression-variables also serve as control-variables.

**Categorical Variables and Fixed-Effects**

In addition to DCF-related control-variables and macroeconomic and macrofinancial variables, several categorical variables may also have deterministic effect on startup valuations. Furthermore, it is likely that at least some of these variables may also serve to modulate the valuation-impact of both macro-level variables and DCF-linked control variables.

Specifically, the nature and scale of the impact of DCF valuation-factors can be influenced by macro-level market conditions, which might be either regulatory or macroeconomic in nature (Gompers and Lerner, 1998; Lockett et al., 2002; Bernoth and Colavecchio, 2014), or by industry-level effects (Chan, 1983; Manigart et al., 1997, 2000; Damodaran, 2009), or by local-level, investor-side, or supply-chain effects (Porter, 1990). Additionally, combinations of these factors impacting the relationship between startup-valuation and valuation-drivers have also been observed for VCs across different countries focusing on fintech-industry startups. Therefore, their inclusion is justified. These include firm, industry, country, city, year, and investor-type drawn primarily from EIKON deals data, as well as from secondary sources such as UK Companies House, which corroborates UK firm-level data, while also contributing geographic data to a substantial level of detail.

# 4. Direct Regression Results

This section demonstrates DCF-consistency, such that hypotheses can viably be tested. Subsequently, we describe direct regression results business-cycle on both startup-valuation and cash-on-market, which test hypotheses H1a, H1b, and H2. Finally, we deploy fixed-effects regressions examining the deterministic role of relevant categorical variables, which test hypothesis 3.

To verify dataset-soundness, we employ a regression-approach inspired by discounted-cashflow valuation, wherein startup-valuations are driven by revenues and by risk-adjusted discount-rates. Damodaran (2009) describes discount-rates as being driven by both sector-level risk-characteristics, and country-risk-premiums.

Table 2 outlines valuation-impact of revenues, sector unlevered-beta, and country-risk-premium, demonstrating that each is statistically-significant, both individually and jointly. These findings both match theoretical-expectations and demonstrate the dataset's theoretical-consistency, with revenue having positive-impact while both beta and country-risk-premium having negative-impact. Goodness-of-fit indicators demonstrate log of revenue as the DCF-driver with strongest explanatory-power, while p-values demonstrate log of sectoral-beta to have the weakest explanatory-power.

*Table 3: Baseline Valuation-Model*

**DCF-based Regressions**

| VARIABLES | (1) Ln_Valuation | (2) Ln_Valuation | (3) Ln_Valuation | (4) Ln_Valuation | (5) Ln_Valuation | (6) Ln_Valuation |
|---|---|---|---|---|---|---|
| Ln_Revenue | 0.6861*** [0.034] | | | 0.6660*** [0.034] | 0.6692*** [0.034] | 0.6514*** [0.034] |
| Ln_Beta | | -2.7738*** [0.395] | | -1.1629*** [0.389] | | -1.0886*** [0.388] |
| Country Risk Premium | | | -70.7055*** [11.701] | | -40.7742*** [13.134] | -38.3685*** [13.092] |
| Constant | 6.4565*** [0.503] | 17.8453*** [0.254] | 16.7791*** [0.133] | 7.4278*** [0.596] | 7.0425*** [0.534] | 7.9171*** [0.615] |
| Observations | 646 | 1,045 | 1,045 | 646 | 646 | 646 |
| R-squared | 0.39 | 0.05 | 0.03 | 0.40 | 0.40 | 0.41 |
| Adjusted R-squared | 0.393 | 0.0442 | 0.0329 | 0.401 | 0.401 | 0.408 |

Standard errors in brackets
*** p<0.01, ** p<0.05, * p<0.1

Nevertheless, generally low R-squared and consistent statistical-significance of the intercept-term indicate that DCF-based regression-model captures only part of the startup-valuation dynamic.

**Direct Business-Cycle Impacts on Startup-Valuation**

We examine the valuation-impact of country-level macro market-conditions, deploying DCF-drivers used in the baseline model as control factors. While empirical literature often treats macro-level market-condition indicators as control-factors (Armstrong et al., 2006; Hellmann and Thiele, 2015), classical approaches such as DCF-based approaches make use of macroeconomic conditions as key valuation-drivers.

Table 3 demonstrates direct valuation-impacts of tax-rates, macroeconomic output-gap, domestic VC cash-on-market, and global VC cash-on-market. Compared to the baseline-model, these regressions grant increased insight, with all regressions expanding goodness-of-fit

The most-deterministic direct macro-level driver is VC cash-on-market, whose inclusion in regressions elevates R-squared to 0.43 for country-level cash-on-market, and 0.53 for global VC cash-on-market, although coefficients for both are extremely-small. Explanatory-power of output gap is also substantial at an R-squared of 0.42. Addition of Dry powder to the regression also elevates R-squared to 0.42, despite boasting an extremely small coefficient, which would indicate that an extremely substantial amount of dry powder would be required to influence relatively-small increases in valuation.

Tax-rates meanwhile, are counterintuitively positive and significant, which may indicate causality flowing from tax-rates rather than towards startup-valuation. Overall, this indicated that.

*Table 4: Direct Impact of Macro-level Startup Valuation-Drivers*

**Macro-based Regressions**

| VARIABLES | (1) Ln_Valuation | (2) Ln_Valuation | (3) Ln_Valuation | (4) Ln_Valuation | (5) Ln_Valuation | (6) Ln_Valuation |
|---|---|---|---|---|---|---|
| Ln_Revenue | 0.6514*** | 0.6540*** | 0.6512*** | 0.7060*** | 0.6862*** | 0.6071*** |
|  | [0.034] | [0.034] | [0.035] | [0.037] | [0.036] | [0.034] |
| Ln_Beta | -1.0886*** | -1.3509*** | -1.1753*** | -0.8804** | -0.9448** | -0.9861*** |
|  | [0.388] | [0.383] | [0.394] | [0.406] | [0.390] | [0.374] |
| Country-Risk-Premium | -38.3685*** | -15.3681 | -13.3416 | -43.2772*** | -23.8422* | -41.5507*** |
|  | [13.092] | [13.622] | [15.048] | [15.708] | [14.386] | [13.232] |
| Tax Rate |  | 0.1130*** |  |  |  |  |
|  |  | [0.020] |  |  |  |  |
| Business-cycle |  |  | 0.2101*** |  |  |  |
|  |  |  | [0.070] |  |  |  |
| Dry-Powder |  |  |  | 0.00000009** |  |  |
|  |  |  |  | [0.000] |  |  |
| Cash-on-Market |  |  |  |  | 0.0006*** |  |
|  |  |  |  |  | [0.000] |  |
| World-Cash-on-Market |  |  |  |  |  | 0.00003*** |
|  |  |  |  |  |  | [0.000] |
| Constant | 7.9171*** | 3.8202*** | 7.9414*** | 6.6447*** | 6.4131*** | 5.3569*** |
|  | [0.615] | [0.933] | [0.634] | [0.687] | [0.648] | [0.639] |
| Observations | 646 | 636 | 632 | 557 | 578 | 550 |
| R-squared | 0.41 | 0.44 | 0.42 | 0.45 | 0.48 | 0.53 |
| Adjusted R-squared | 0.408 | 0.438 | 0.421 | 0.447 | 0.473 | 0.526 |

Standard errors in brackets
*** p<0.01, ** p<0.05, * p<0.1

Examination of direct-effects on valuation also demands examination of business-cycle-interaction-effects. Table 4 examines interaction-effects between DCF-factors and macroeconomic factors on valuation.

Panel A examines DCF-output-gap interaction-effects in addition to output-gap, demonstrating output-gaps to have significant interaction-effects on risk-adjusted financing costs, acting on both country-risk-premium and sectoral-beta. Panel B, which includes DCF-dry-powder interaction effects, finds that dry powder only has (very slight) interaction-effects with country-risk-premium. Meanwhile, Panel C examines DCF-cash-on-market interaction-effects, demonstrating cash-on-market to mainly have interaction-effects with country-risk-premium. This indicates that cash-on-market likely plays a larger role in determining country-risk-premium's impact on startup-valuation. Meanwhile, Panel D demonstrates interaction-effects of world-cash-on-market to be negligible. This means that interaction-effects are driven by domestic economic-trends rather than by international ones.

Compared to Table 3, inclusion of world cash-on-market interaction-effects improves goodness-of-fit figures, while inclusion of output-gap-interaction-effects and domestic cash-on-market have either no effect on goodness-of-fit figures, or slightly deceased them.

*Table 5: Macroeconomic Interaction Direct-effects*

**Panel A: Output-Gap Interaction Effects**

| VARIABLES | (1) Ln_Valuation | (2) Ln_Valuation | (3) Ln_Valuation | (4) Ln_Valuation |
|---|---|---|---|---|
| Ln_Revenue | 0.6480*** | 0.6498*** | 0.6390*** | 0.6470*** |
|  | [0.040] | [0.041] | [0.040] | [0.040] |
| Ln_Beta | -0.9043** | -0.9022** | -0.9306** | -1.2815*** |
|  | [0.440] | [0.440] | [0.439] | [0.490] |
| Country-Risk-Premium | -17.8211 | -17.7684 | -29.7066 | -15.5717 |
|  | [17.098] | [17.115] | [18.470] | [17.116] |
| Output Gap | 0.2715*** | 0.2693*** | 0.4149*** | 0.6988*** |
|  | [0.081] | [0.082] | [0.117] | [0.260] |
| RevenuexOutput Gap |  | 0.0000 |  |  |
|  |  | [0.000] |  |  |
| Country-Risk-PremiumxOutput Gap |  |  | -9.6545* |  |
|  |  |  | [5.733] |  |
| BetaxOutput Gap |  |  |  | -0.4522c |
|  |  |  |  | [0.262] |
| Constant | 7.9412*** | 7.9145*** | 8.1827*** | 8.1886*** |
|  | [0.702] | [0.714] | [0.715] | [0.715] |
| Observations | 547 | 547 | 547 | 547 |
| R-squared | 0.41 | 0.41 | 0.41 | 0.41 |
| Adjusted R-squared | 0.402 | 0.401 | 0.404 | 0.404 |

**Panel B: Dry Powder Interaction Regressions**

| VARIABLES | (1) Ln_Valuation | (2) Ln_Valuation | (3) Ln_Valuation | (4) Ln_Valuation |
|---|---|---|---|---|
| Ln_Revenue | 0.7060*** | 0.7297*** | 0.6691*** | 0.7055*** |
|  | [0.037] | [0.051] | [0.037] | [0.037] |
| Ln_Beta | -0.8804** | -0.5114 | -1.0815*** | -0.9664 |
|  | [0.406] | [0.455] | [0.397] | [0.730] |
| Country-Risk-Premium | -43.2772*** | -52.4704*** | 48.6125** | -43.0770*** |
|  | [15.708] | [17.187] | [22.685] | [15.786] |
| Dry Powder | 0.00000009** | 0.0000001*** | 0.0000006*** | 0.0000 |
|  | [0.000] | [0.000] | [0.000] | [0.000] |
| Revenue x Dry Powder |  | -0.0000 |  |  |
|  |  | [0.000] |  |  |
| Country-Risk-Premium x Dry Powder |  |  | -0.0001*** |  |
|  |  |  | [0.000] |  |
| Beta x Dry Powder |  |  |  | 0.0000 |
|  |  |  |  | [0.000] |
| Constant | 6.6447*** | 6.1210*** | 6.1716*** | 6.6993*** |
|  | [0.687] | [0.857] | [0.675] | [0.788] |
| Observations | 557 | 494 | 557 | 557 |
| R-squared | 0.45 | 0.43 | 0.48 | 0.45 |
| Adjusted R-squared | 0.447 | 0.421 | 0.475 | 0.446 |

Standard errors in brackets
*** p<0.01, ** p<0.05, * p<0.1

**Panel C: Cash-on-Market Interaction Effects**

| VARIABLES | (1) Ln_Valuation | (2) Ln_Valuation | (3) Ln_Valuation | (4) Ln_Valuation |
|---|---|---|---|---|
| Ln_Revenue | 0.6859*** | 0.6824*** | 0.6298*** | 0.6583*** |
|  | [0.039] | [0.051] | [0.042] | [0.042] |
| Ln_Beta | -0.5289 | -0.4084 | -0.6685 | 1.9392*** |

|  | [0.430] | [0.479] | [0.467] | [0.729] |
| --- | --- | --- | --- | --- |
| Country-Risk-Premium | -33.4035** | -42.9594** | 49.1285* | -40.5828** |
|  | [15.701] | [20.361] | [25.972] | [19.785] |
| Cash-on-Market | 0.0008*** | 0.0007*** | 0.0016*** | 0.0014*** |
|  | [0.000] | [0.000] | [0.000] | [0.000] |
| RevenuexCash-on-Market |  | -0.0000 |  |  |
|  |  | [0.000] |  |  |
| Country-Risk-PremiumxCash-on-Market |  |  | -0.1306*** |  |
|  |  |  | [0.024] |  |
| BetaxCash-on-Market |  |  |  | -0.0012*** |
|  |  |  |  | [0.000] |
| Constant | 6.1254*** | 6.1676*** | 6.4098*** | 5.3686*** |
|  | [0.703] | [0.841] | [0.749] | [0.784] |
| Observations | 514 | 455 | 455 | 455 |
| R-squared | 0.47 | 0.43 | 0.46 | 0.45 |
| Adjusted R-squared | 0.461 | 0.424 | 0.458 | 0.446 |

**Panel D: World Cash-on-Market Interaction Effects**

| VARIABLES | (1) Ln_Valuation | (2) Ln_Valuation | (3) Ln_Valuation | (4) Ln_Valuation |
| --- | --- | --- | --- | --- |
| Ln_Revenue | 0.5917*** | 0.5586*** | 0.5819*** | 0.5944*** |
|  | [0.038] | [0.047] | [0.038] | [0.039] |
| Ln_Beta | -0.6105 | -0.6690 | -0.6581 | 0.0246 |
|  | [0.413] | [0.416] | [0.412] | [1.480] |
| Country-Risk-Premium | -55.1308*** | -53.6501*** | -166.5205*** | -54.1948*** |
|  | [14.513] | [14.555] | [47.848] | [14.676] |
| World Cash-on-Market | .000032*** | .000036*** | .000028*** | .000036*** |
|  | [0.000] | [0.000] | [0.000] | [0.000] |
| RevenuexWorld Cash |  | 0.0000 |  |  |
|  |  | [0.000] |  |  |
| Country-Risk-PremiumxWorld Cash |  |  | 0.0010** |  |
|  |  |  | [0.000] |  |
| BetaxWorld Cash |  |  |  | -0.0000 |
|  |  |  |  | [0.000] |
| Constant | 5.2088*** | 5.5788*** | 6.2778*** | 4.7879*** |
|  | [0.689] | [0.751] | [0.813] | [1.167] |
| Observations | 486 | 486 | 486 | 486 |
| R-squared | 0.52 | 0.52 | 0.52 | 0.52 |
| Adjusted R-squared | 0.514 | 0.514 | 0.519 | 0.513 |

Standard errors in brackets
a p<0.01, b p<0.05, c p<0.1

These findings demonstrate evidence confirming the direct-effects on startup-valuations outlined by Hypothesis 1a and 1b. Goodness-of-fit figures outlined in Tables 3 and 4 demonstrate substantial increases in goodness-of-fit, which can be taken as evidence of direct-effect valuation-impacts.

## Cash-on-Market

In addition to the valuation-impact of drivers on startups, business-cycle might also drive cash-on-market. Because cash-on-market demonstrably plays a deterministic role in pre-money startup valuation, impacts macro-level on cash-on-market can be considered *second-order effects*.

Table 5 examines impacts of macro-level factors on domestic cash-on-market, with Panel A examining univariate-regressions and Panel B multivariate-regressions. Overall, goodness-of-fit indicators demonstrate that output-gap and world cash-on-market have the strongest explanatory power. This is followed by country-risk-premium. Both tax rates and Country-risk-Premiums have negative impacts on domestic cash-on-market, in line with economic theory.

Panel B regressions combining these drivers indicate that while business-cycle and funding-costs drive domestic-cash-on-market, inclusion of world cash-on-market more than doubles the regression's goodness-of-fit, indicating strong evidence of cross-border impact on cash-on-market.

*Table 6: Determinants of Cash-on-Market*

**Panel A: Cash-on-Market Regressions**

| VARIABLES | (1) Cash-on-Market | (2) Cash-on-Market | (3) Cash-on-Market | (4) Cash-on-Market |
|---|---|---|---|---|
| Output Gap | 165.2337*** | | | |
|  | [17.563] | | | |
| Tax-Rate | | -20.0792*** | | |
|  | | [5.848] | | |
| Country-Risk-Premium | | | -25,101.1220*** | |
|  | | | [4,057.279] | |
| Cash on World Market | | | | 0.0148*** |
|  | | | | [0.001] |
| Constant | 1,362.4216*** | 1,943.5066*** | 1,438.5839*** | -210.4132** |
|  | [30.122] | [209.957] | [44.071] | [83.367] |
| Observations | 800 | 810 | 816 | 787 |
| R-squared | 0.10 | 0.01 | 0.04 | 0.27 |
| Adjusted R-squared | 0.0987 | 0.0132 | 0.0437 | 0.272 |

**Panel B: Joint Cash-on-Market Regressions**

| VARIABLES | (1) Cash-on-Market | (2) Cash-on-Market | (3) Cash-on-Market | (4) Cash-on-Market | (5) Cash-on-Market | (6) Cash-on-Market |
|---|---|---|---|---|---|---|
| Output Gap | 174.6744*** | 132.9126*** | 111.6870*** | 105.3422*** | 54.5474*** | 45.2741** |
|  | [17.420] | [20.934] | [14.746] | [14.872] | [18.119] | [17.700] |
| Tax-Rate | -23.8595*** | -28.1054*** | -15.4146*** | | | -22.0115*** |
|  | [5.681] | [5.767] | [4.630] | | | [4.627] |
| Country-Risk-Premium | | -17,418.2210*** | | | -18,805.5999*** | -25,471.4445*** |
|  | | [4,919.879] | | | [3,939.687] | [3,962.853] |
| Cash on World Market | | | 0.0126*** | 0.0127*** | 0.0134*** | 0.0135*** |
|  | | | [0.001] | [0.001] | [0.001] | [0.001] |
| Constant | 2,224.7143*** | 2,489.1710*** | 646.5089*** | 76.6646 | 129.0050 | 957.9061*** |
|  | [205.521] | [217.284] | [189.237] | [89.485] | [88.921] | [190.679] |
| Observations | 795 | 795 | 766 | 771 | 771 | 766 |
| R-squared | 0.12 | 0.14 | 0.33 | 0.31 | 0.33 | 0.36 |
| Adjusted R-squared | 0.119 | 0.132 | 0.326 | 0.312 | 0.331 | 0.360 |

Standard errors in brackets
a $p<0.01$, b $p<0.05$, c $p<0.1$

Overall, this demonstrates that Hypothesis 2 is established, given that within our dataset, domestic cash-on-market appears to be driven by business-cycle indicators, as well as tax-rates and country-risk-premiums.

Essentially, this establishes that macro-level market-conditions have a *two-channel-impact* on startup-valuations, meaning that the total market-condition-effect may be slightly more complex than described by Berre and Le Pendeven (2022). These results are corroborated by multicollinearity testing, establishing the independence of each of these valuation channels from one-another, outlined in Appendix II.

In addition to macro-level drivers, our dataset includes several categorical-variables capturing a range of localized small-scale-effects. These include variables for firm, sector, investor-type, city, and country.

Table 6 includes fixed-effects regressions examining direct valuation-impacts outlined in Table 3. Panel A conveys firm-fixed-effects, while Panel B conveys sector-fixed effects, Panel C outlines investor-type fixed-effects, Panel D city fixed-effects, and Panel E year fixed-effects. In terms of overall goodness-of-fit, sector, year, and fixed-effects models capture the largest amounts of variation in startup-valuation within our dataset, while firm fixed-effects capture the least variation in startup valuation.

Meanwhile, macro-regression between-group r-squared is particularly large for year fixed-effects and city fixed-effects, with between-group r-squared demonstrating the majority explanatory-effect for investor-type, city, and year fixed-effects, while within-group r-squared demonstrates the majority of explanatory-power for industry fixed-effects.

Because all panels include the baseline DCF model as regression 1, as well as regressions for both output-gap and cash-on-market, we directly compare effects of the categorical-variables on the explanatory-power of DCF-factors to their effects on explanatory-power of macro-level factors. Differences between goodness-of-fit indicators between Table 3 and Table 5 indicate that while the explanatory-power added-value of our fixed-effects models is concentrated in its effect on macro-level indicators.

In terms of driver-impact meanwhile, domestic cash-on-world-market is the most consistently-influential valuation-driver. Country-risk-premium however loses its statistical-significance in all panels, indicating that our categorical replicate at least some of the descriptive-power expressed country-risk-premia.

*Table 7: Direct-Effect FE Regressions*

**Fixed-Effects Macro-Regressions Panel A: Firm Fixed-Effects**

| VARIABLES | (1) Ln_Valuation | (2) Ln_Valuation | (3) Ln_Valuation | (4) Ln_Valuation | (5) Ln_Valuation | (6) Ln_Valuation |
|---|---|---|---|---|---|---|
| Ln_Revenue | 0.4878*** | 0.4704*** | 0.5678*** | 0.5611*** | 0.5307*** | 0.5102*** |
|  | [0.044] | [0.043] | [0.045] | [0.049] | [0.048] | [0.047] |
| Ln_Beta | -1.6991*** | -1.9726*** | -1.0180* | -1.4501** | -1.6331*** | -1.6174*** |
|  | [0.542] | [0.527] | [0.540] | [0.568] | [0.559] | [0.546] |
| Country-Risk-Premium | -66.0262*** | -58.8327*** | 12.6377 | -84.4909*** | -72.5969*** | -69.7739*** |
|  | [10.093] | [10.002] | [16.880] | [11.659] | [11.151] | [10.860] |
| Tax-Rates |  | 0.1232*** |  |  |  |  |
|  |  | [0.025] |  |  |  |  |
| Output-Gap |  |  | 0.3408*** |  |  |  |
|  |  |  | [0.059] |  |  |  |
| Dry-Powder |  |  |  | 0.0000 |  |  |
|  |  |  |  | [0.000] |  |  |
| Cash-on-Market |  |  |  |  | 0.0000 |  |
|  |  |  |  |  | [0.000] |  |
| Cash on World Market |  |  |  |  |  | 0.0000 |
|  |  |  |  |  |  | [0.000] |
| Constant | 10.3814*** | 6.3451*** | 8.3925*** | 9.1870*** | 9.6259*** | 9.8025*** |
|  | [0.624] | [1.012] | [0.707] | [0.705] | [0.685] | [0.668] |
| Observations | 646 | 636 | 632 | 557 | 578 | 550 |
| chi2 | 153.7 | 187.1 | 191.4 | 161.9 | 155.2 | 149.7 |
| Within R2 | 0.0551 | 0.0551 | 0.0551 | 0.0549 | 0.0496 | 0.0495 |
| Between R2 | 0.293 | 0.343 | 0.350 | 0.349 | 0.331 | 0.321 |
| Overall R2 | 0.394 | 0.413 | 0.418 | 0.431 | 0.433 | 0.440 |

Standard errors in brackets
\*\*\* p<0.01, \*\* p<0.05, \* p<0.1

**Fixed-Effects Macro-Regressions: Panel B: Sector Fixed-Effects**

| VARIABLES | (1) Ln_Valuation | (2) Ln_Valuation | (3) Ln_Valuation | (4) Ln_Valuation | (5) Ln_Valuation | (6) Ln_Valuation |
|---|---|---|---|---|---|---|
| Ln_Revenue | 0.6318*** | 0.6367*** | 0.6397*** | 0.6943*** | 0.6787*** | 0.6091*** |
|  | [0.034] | [0.034] | [0.035] | [0.037] | [0.036] | [0.034] |
| Ln_Beta | -0.3805 | -0.8452 | -0.4637 | 0.7676 | 0.8712 | 1.3837** |
|  | [0.578] | [0.528] | [0.584] | [0.717] | [0.702] | [0.700] |
| Country-Risk-Premium | -38.3666*** | -15.7142 | -18.1501 | -40.7037*** | -25.0954* | -40.9948*** |
|  | [12.737] | [13.328] | [14.747] | [14.972] | [13.765] | [12.543] |
| Tax-Rates |  | 0.1089*** |  |  |  |  |
|  |  | [0.019] |  |  |  |  |

| VARIABLES | (1) | (2) | (3) | (4) | (5) | (6) |
|---|---|---|---|---|---|---|
| Output-Gap | | | 0.1604** | | | |
| | | | [0.069] | | | |
| Dry-Powder | | | | 0.0000000765* | | |
| | | | | [0.000] | | |
| Cash-on-Market | | | | | 0.0006*** | |
| | | | | | [0.000] | |
| Cash on World Market | | | | | | 0.0000*** |
| | | | | | | [0.000] |
| Constant | 7.6960*** | 3.8480*** | 7.5952*** | 5.7472*** | 5.4232*** | 3.8673*** |
| | [0.690] | [0.951] | [0.709] | [0.814] | [0.779] | [0.776] |
| Observations | 643 | 633 | 630 | 554 | 575 | 547 |
| chi2 | 395.5 | 452.4 | 406.0 | 397.9 | 450.9 | 554.0 |
| Within R2 | 0.356 | 0.387 | 0.366 | 0.406 | 0.429 | 0.501 |
| Between R2 | 0.691 | 0.720 | 0.705 | 0.654 | 0.681 | 0.685 |
| Overall R2 | 0.409 | 0.443 | 0.423 | 0.435 | 0.457 | 0.493 |

Standard errors in brackets
*** p<0.01, ** p<0.05, * p<0.1

**Fixed-Effects Macro-Regressions: Panel C: Investor-Type Fixed-Effects**

| VARIABLES | (1) Ln_Valuation | (2) Ln_Valuation | (3) Ln_Valuation | (4) Ln_Valuation | (5) Ln_Valuation | (6) Ln_Valuation |
|---|---|---|---|---|---|---|
| Ln_Revenue | 0.4049*** | 0.4173*** | 0.4192*** | 0.4268*** | 0.4252*** | 0.4073*** |
| | [0.034] | [0.035] | [0.035] | [0.036] | [0.036] | [0.036] |
| Ln_Beta | -1.1420*** | -1.2143*** | -1.0146*** | -0.8366** | -0.7996** | -0.7228** |
| | [0.363] | [0.367] | [0.367] | [0.364] | [0.361] | [0.361] |
| Country-Risk-Premium | -24.6074* | -11.3366 | -12.5279 | -20.3562 | -17.4586 | -26.9445** |
| | [12.611] | [13.439] | [14.542] | [14.702] | [13.912] | [13.556] |
| Tax-Rates | | 0.0568*** | | | | |
| | | [0.022] | | | | |
| Output-Gap | | | 0.0419 | | | |
| | | | [0.069] | | | |
| Dry-Powder | | | | 0.0000000677* | | |
| | | | | [0.000] | | |
| Cash-on-Market | | | | | 0.0003** | |
| | | | | | [0.000] | |
| Cash on World Market | | | | | | 0.0000*** |
| | | | | | | [0.000] |
| Constant | 11.4272*** | 9.1675*** | 11.1241*** | 10.7342*** | 10.6453*** | 9.1087*** |
| | [0.693] | [1.065] | [0.700] | [0.759] | [0.748] | [0.757] |
| Observations | 528 | 518 | 514 | 460 | 467 | 447 |
| chi2 | 185.8 | 194.7 | 188.3 | 173.9 | 182.8 | 216.8 |
| Within R2 | 0.253 | 0.264 | 0.260 | 0.264 | 0.271 | 0.295 |
| Between R2 | 0.619 | 0.654 | 0.653 | 0.640 | 0.673 | 0.769 |
| Overall R2 | 0.399 | 0.434 | 0.409 | 0.436 | 0.451 | 0.495 |

Standard errors in brackets
*** p<0.01, ** p<0.05, * p<0.1

**Fixed-Effects Macro-Regressions: Panel D: City Fixed-Effects**

| VARIABLES | (1) Ln_Valuation | (2) Ln_Valuation | (3) Ln_Valuation | (4) Ln_Valuation | (5) Ln_Valuation | (6) Ln_Valuation |
|---|---|---|---|---|---|---|
| Ln_Revenue | 0.6022*** | 0.6084*** | 0.5958*** | 0.5756*** | 0.5834*** | 0.5615*** |
| | [0.037] | [0.037] | [0.037] | [0.037] | [0.035] | [0.035] |
| Ln_Beta | -1.3550*** | -1.3592*** | -1.4484*** | -1.5668*** | -1.4295*** | -1.3792*** |
| | [0.401] | [0.401] | [0.402] | [0.379] | [0.365] | [0.371] |
| Country-Risk-Premium | -43.0762* | -20.8989 | -12.2993 | -16.7040 | -1.9419 | -56.0570** |
| | [23.707] | [24.087] | [25.540] | [30.958] | [25.205] | [23.186] |
| Tax-Rates | | 0.0912** | | | | |
| | | [0.036] | | | | |
| Output-Gap | | | 0.3019*** | | | |
| | | | [0.089] | | | |
| Dry-Powder | | | | 0.0000000677*** | | |
| | | | | [0.000] | | |
| Cash-on-Market | | | | | 0.0015*** | |
| | | | | | [0.000] | |
| Cash on World Market | | | | | | 0.0000*** |
| | | | | | | [0.000] |
| Constant | 9.0680*** | 5.4461*** | 9.2576*** | 7.7852*** | 7.7240*** | 6.5933*** |
| | [0.695] | [1.533] | [0.703] | [0.696] | [0.663] | [0.700] |
| Observations | 588 | 580 | 576 | 519 | 539 | 511 |
| chi2 | 302.7 | 312.1 | 319.0 | 467.3 | 534.4 | 439.6 |
| Within R2 | 0.325 | 0.324 | 0.336 | 0.518 | 0.533 | 0.438 |
| Between R2 | 0.410 | 0.464 | 0.473 | 0.180 | 0.273 | 0.605 |
| Overall R2 | 0.418 | 0.445 | 0.427 | 0.328 | 0.420 | 0.532 |

Standard errors in brackets

**Fixed-Effects Macro-Regressions: Panel E: Year Fixed-Effects**

| VARIABLES | (1) Ln_Valuation | (2) Ln_Valuation | (3) Ln_Valuation | (4) Ln_Valuation | (5) Ln_Valuation | (6) Ln_Valuation |
|---|---|---|---|---|---|---|
| Ln_Revenue | 0.5561*** | 0.5563*** | 0.5678*** | 0.5837*** | 0.5943*** | 0.5857*** |
|  | [0.031] | [0.030] | [0.031] | [0.033] | [0.033] | [0.033] |
| Ln_Beta | -1.3717*** | -1.5432*** | -1.2613*** | -1.1638*** | -1.2216*** | -1.1624*** |
|  | [0.345] | [0.338] | [0.344] | [0.350] | [0.348] | [0.358] |
| Country-Risk-Premium | -41.2498*** | -21.1418* | -49.6215*** | -55.9876*** | -47.8448*** | -41.1296*** |
|  | [11.507] | [12.059] | [13.736] | [13.533] | [12.975] | [12.651] |
| Tax-Rates |  | 0.0911*** |  |  |  |  |
|  |  | [0.018] |  |  |  |  |
| Output-Gap |  |  | -0.1358* |  |  |  |
|  |  |  | [0.072] |  |  |  |
| Dry-Powder |  |  |  | -0.0000*** |  |  |
|  |  |  |  | [0.000] |  |  |
| Cash-on-Market |  |  |  |  | -0.0002* |  |
|  |  |  |  |  | [0.000] |  |
| Cash on World Market |  |  |  |  |  | 0.0000 |
|  |  |  |  |  |  | [0.000] |
| Constant | 9.6748*** | 6.2565*** | 9.5069*** | 9.7472*** | 9.3843*** | 7.8145*** |
|  | [0.644] | [0.930] | [0.672] | [0.735] | [0.708] | [1.056] |
| Observations | 646 | 636 | 632 | 557 | 578 | 550 |
| chi2 | 427.5 | 466.9 | 439.7 | 456.7 | 449.6 | 413.8 |
| Within R2 | 0.409 | 0.433 | 0.423 | 0.459 | 0.446 | 0.432 |
| Between R2 | 0.000880 | 0.00580 | 0.00588 | 0.121 | 0.206 | 0.297 |
| Overall R2 | 0.409 | 0.440 | 0.397 | 0.406 | 0.416 | 0.507 |

Standard errors in brackets
*** p<0.01, ** p<0.05, * p<0.1

Table 7, using joint-fixed-effects, demonstrated that inclusion of joint-fixed-effects renders valuation-impacts of both sectoral-beta and macroeconomic output-gap statistically-insignificant, indicating that categorical-variables used in this joint fixed-effect regression capture similar value-signal information to sectoral-beta and business-cycle-indicators.

Nevertheless, because both cash-on-market and country-risk-premium remain statistically-significant, it is likely that the joint-fixed-effects model's categorical variables capture-and-substitute direct, but not indirect business-cycle-impacts on startup-valuation.

*Table 8: Direct-Effect FE Regressions: Sector, City, Investor-Type*

**Fixed-Effects Macro-Regressions: Joint Fixed-Effects**

| VARIABLES | (1) Ln_Valuation | (2) Ln_Valuation | (3) Ln_Valuation | (4) Ln_Valuation | (5) Ln_Valuation | (6) Ln_Valuation |
|---|---|---|---|---|---|---|
| Ln_Revenue | 0.4962*** | 0.4946*** | 0.4944*** | 0.5014*** | 0.5057*** | 0.4677*** |
|  | [0.039] | [0.039] | [0.040] | [0.041] | [0.040] | [0.040] |
| Ln_Beta | 0.1123 | -0.2099 | 0.0956 | 0.1780 | 0.2403 | 0.0264 |
|  | [0.593] | [0.591] | [0.601] | [0.629] | [0.613] | [0.586] |
| Country-Risk-Premium | -76.3381*** | -37.3387 | -60.0232** | -73.4863** | -58.7941** | -83.6040*** |
|  | [23.359] | [26.552] | [27.346] | [28.933] | [26.648] | [24.629] |
| Tax-Rates |  | 0.1181*** |  |  |  |  |
| Output-Gap | | [0.036] | | 0.1176 | | |
| | | | | [0.094] | | |
| Dry-Powder | | | | | 0.0000000908*** | |
| | | | | | [0.000] | |
| Cash-on-Market | | | | | | 0.0006*** |
| | | | | | | [0.000] |
| Cash on World Market | | | | | | 0.0000*** |
| | | | | | | [0.000] |
| Constant | 9.8603*** | 5.5521*** | 9.8881*** | 9.0741*** | 8.8663*** | 7.6748*** |
| | [0.759] | [1.501] | [0.779] | [0.826] | [0.815] | [0.809] |
| Observations | 474 | 466 | 462 | 425 | 431 | 411 |
| chi2 | 185.7 | 198.6 | 183.1 | 179.6 | 191.2 | 244.0 |
| Within R2 | 0.190 | 0.185 | 0.184 | 0.239 | 0.230 | 0.263 |
| Between R2 | 0.391 | 0.422 | 0.408 | 0.391 | 0.416 | 0.503 |
| Overall R2 | 0.412 | 0.444 | 0.419 | 0.413 | 0.439 | 0.495 |

Standard errors in brackets
a p<0.01, b p<0.05, c p<0.1

Overall, the findings of fixed-effects models are generally-similar to those of macro-based OLS, both in terms of both coefficient-significance and model-goodness of fit. Exception to this are the results of joint-fixed-effects model, which find the model's categorical-variables eclipse the explanatory-power of sectoral-beta and output-gap, which are statistically non-significant in the joint-fixed-effects model.

# 5. Functional-Form and Structural Equation Results

In addition to direct regression-results, we examine functional-form questions and indirect regression-results, these test hypotheses 4a, 4b, and 4c via use of mediation-structure-testing, endogeneity-testing, and structural-equation-model regressions, testing both circular and non-circular functional-forms.

**Functional-Form**

Sobel-testing can be used to establish mediation and partial-mediation structural-relationships. Table 8 examines mediation-structure likelihood between business-cycle and startup-valuation, testing the mediation-role of both domestic cash-on-market and credit-risk-premium, establishing mediation-structure of both cash-on-market and credit-risk-premium.

*Table 9: Sobel testing for mediation*

| | | | | Test statistic: | SE | P-value: |
|---|---|---|---|---|---|---|
| **Sobel: Output-Gap, Cash-on-Market, Valuation** | | | | | | |
| Output-Gap and Cash-on-Market | Coefficient | 165.23370 | Sobel test: | 5.91092 | 0.02825 | 0.000*** |
| | SE | 17.56309 | Aroian test: | 5.89081 | 0.02834 | 0.000*** |
| Cash-on-Market and Valuation | Coefficient | 0.00101 | Goodman test: | 5.93123 | 0.02815 | 0.000*** |
| | SE | 0.00013 | | | | |
| **Sobel: Output-Gap, Country-Risk Premium, Valuation** | | | | Test statistic: | SE | P-value: |
| Output-Gap and Country-Risk Premium | Coefficient | -0.00190 | Sobel test: | 5.60841 | 0.02399 | 0.000*** |
| | Standard Error | 0.00013 | Aroian test: | 5.59798 | 0.02403 | 0.000*** |
| Country-Risk Premium and Valuation | Coefficient | -70.69590 | Goodman test: | 5.61890 | 0.02394 | 0.000*** |
| | Standard Error | 11.71862 | | | | |
| **Sobel: Output-Gap, Dry Powder, Valuation** | | | | Test statistic: | SE | P-value: |
| Output-Gap and Dry Powder | Coefficient | 313042.60000 | Sobel test: | 1.66041 | 0.01623 | 0.09683148* |
| | Standard Error | 52544.28000 | Aroian test: | 1.63925 | 0.01644 | 0.10116 |
| Dry Powder and Valuation | Coefficient | 0.0000000861 | Goodman test: | 1.68242 | 0.01602 | 0.09248829* |
| | Standard Error | 0.0000000498 | | | | |

## Endogeneity

Execution of a VAR model, accompanied by impulse-response-function, and optimal-lag-selection testing examine endogenous-relationship likelihood, impact, and impact-decomposition by looking-into possibilities for circular-causality.

Log-transformation of all dependent and independent variables gives us a scaled, readable impulse-response-function. Figure 2 demonstrates the impulse-response function of all log-transformed variables, outlining that the majority of the macro-level effects can be manifest primarily after a five yearly-period delay.

*Figure 2: Impulse Response Functions: Output Gap, Cash-on-Market, Valuation*

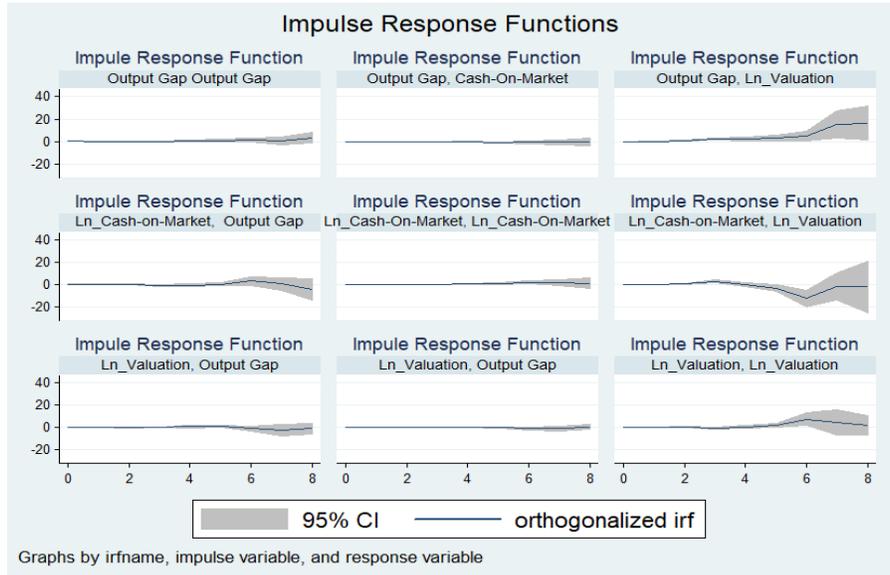

This long time-horizon effect is corroborated by Table 9, which displays optimal-lag testing, finding that that the optimal number of yearly-lags be set at seven years. While it is indeed possible that startup-valuations may give-rise to endogeneity via circular-causality, by attracting increased domestic cash-on-market, the relationship between valuations and cash-on market is demonstrated to be extremely slow-moving.

*Table 10: VAR Model Optimal Lag Selection*

| Selection-order | criteria | | | | | Observations | 53 | |
|---|---|---|---|---|---|---|---|---|
| Sample: | 26 | -1068 | but with gaps | | | | | |
| lag | LL | LR | df | p | FPE | AIC | HQIC | SBIC |
| 0 | -595.428 | | | | 2.00E+06 | 23.035 | 23.2495 | 23.5926 |
| 1 | -523.693 | 143.47 | 9 | 0.000 | 191186 | 20.6677 | 21.0108 | 21.5599 |
| 2 | -512.626 | 22.133 | 9 | 0.008 | 178868 | 20.5897 | 21.0614 | 21.8165 |
| 3 | -492.2 | 40.852 | 9 | 0.000 | 118582 | 20.1585 | 20.7589 | 21.7199 |
| 4 | -479.089 | 26.222 | 9 | 0.002 | 104834 | 20.0034 | 20.7325 | 21.8993 |
| 5 | -451.069 | 56.04 | 9 | 0.000 | 53621.4 | 19.2856 | 20.1434 | 21.5161* |
| 6 | -438.66 | 24.819 | 9 | 0.003 | 50421.2 | 19.157 | 20.1434 | 21.7221 |
| 7 | -420.999 | 35.32* | 9 | 0.000 | 39893.5* | 18.8302* | 19.9452* | 21.7298 |

The optimal-lag-test findings corroborate Sedláček and Sterk (2017), which outline that US startups are influenced by macro-level and business-cycle conditions at the time of their entry, with effects persisting for years. These findings are also in-line with ECB (2021), which outlines that non-bank financial intermediaries such as VC funds are relatively more responsive to changes in long-term interest rates.

**Structural Equation Approach**

Table 10 estimates the relationship between startup-valuation, output-gap and cash-on-market, as a dual-partial-mediation relationship described in Figure 1A and in Equation 2. The findings indicate that startup-valuation is directly-impacted by output gap, country-risk-premium, and cash-on-market, and also indirectly-impacted, as country-risk-premium and cash-on-market are impacted by output-gap. Goodness-of-fit indicators indicate that the total model has substantially-improved explanatory-power compared to any direct-effect OLS models or fixed-effects models, and that indirect-effects via cash-on-market play a more-prominent role in explaining startup-valuation than indirect-effects via country-risk-premiums do. This finding corroborates Table 5 findings, in which domestic cash-on-market is the most consistently-significant valuation-driver, while country-risk-premium loses statistical-significance in some regressions in each panel of Table 5.

*Table 11: Structural Equation Model Regression*

| Structural equation model<br>Estimation method = ml<br>Log likelihood = -8064.2694 | Observations: | 1089 | | | | |
|---|---|---|---|---|---|---|
| | Coef. | SE | z | P-Value | [95% Conf. | Interval] |
| Dependent: Ln_Valuation | | | | | | |
| Cash-on-Market | 0.0006 | 0.0001 | 5.94 | 0.000*** | 0.0004 | 0.0008 |
| Country-Risk-Premium | -9.3936 | 11.1616 | -0.84 | 0.4000 | -31.2700 | 12.4828 |
| Output-Gap | 0.1805 | 0.0573 | 3.15 | 0.002*** | 0.0683 | 0.2928 |
| Ln_Revenue | 0.5708 | 0.0295 | 19.34 | 0.000*** | 0.5130 | 0.6287 |
| Ln_Beta | -1.5794 | 0.3336 | -4.73 | 0.000*** | -2.2332 | -0.9257 |
| Constant | 8.2520 | 0.5412 | 15.25 | 0.000*** | 7.1912 | 9.3127 |
| Dependent: Cash-on-Market | | | | | | |
| Country-Risk-Premium | -5257.4110 | 4748.7620 | -1.11 | 0.2680 | -14564.8100 | 4049.9910 |
| Output-Gap | 153.6415 | 21.5621 | 7.13 | 0.000*** | 111.3807 | 195.9024 |
| Constant | 1405.4680 | 42.7423 | 32.88 | 0.000*** | 1321.6950 | 1489.2410 |
| Dependent: Country-Risk-Premium | | | | | | |
| Output-Gap | -0.0021 | 0.0001 | -15.53 | 0.000*** | -0.0024 | -0.0018 |
| Constant | 0.0077 | 0.0002 | 37.03 | 0.000*** | 0.0073 | 0.0081 |
| | chi2(7) | 21.23 | | | Prob. > chi2 | 0.0117 |

| Wald tests for equations | | | |
|---|---|---|---|
| observed | chi2 | df | p |
| Ln_Valuation | 530.19 | 5 | 0.000*** |
| Cash-on-Market | 242.74 | 1 | 0.000*** |
| Credit-Risk-Premium | 90.05 | 2 | 0.000*** |

| Equation-level goodness-of-fit | | | | | | |
|---|---|---|---|---|---|---|
| Dependent Variable | Fitted Variance | Predicted Variance | Residual Variance | R-Squared | mc | mc2 |
| Ln_Valuation | 7.592008 | 3.077912 | 4.514096 | 0.4054148 | 0.6367219 | 0.4054148 |
| Cash-on-Market | 0.0000503 | 9.86E-06 | 0.0000405 | 0.195894 | 0.4425992 | 0.195894 |
| Credit-Risk-Premium | 609226.6 | 62144.88 | 547081.7 | 0.1020062 | 0.3193841 | 0.1020062 |
| Overall | | | | 0.5240205 | | |

Table 11 estimates the relationship between startup-valuation, business-cycle-indicators and cash-on-market, as a dual-partial-mediation relationship structured to capture circular-causality, as per Figure 1B and Equation 3, finding that while startup-valuation is indeed impacted by output gap, country-risk-premium, and cash-on-market, and that country-risk-premium and cash-on-market are impacted by output-gap. Meanwhile, cash-on-market is not significantly-impacted by startup-valuation.

*Table 12: Structural-Equation Model Regression Partial Circular-Causality*

Structural equation model
Estimation method = ml   Observations: 514
Log likelihood = -5469.6096

| | Coef. | SE | Z | P-Value | [95% Conf. | Interval] |
|---|---|---|---|---|---|---|
| Dependent: Ln_Valuation | | | | | | |
| Cash-on-Market | 0.0005 | 0.0002 | 2.57 | 0.010*** | 0.0001 | 0.0009 |
| Country-Risk-Premium | -17.6720 | 17.7988 | -0.99 | 0.3210 | -52.5570 | 17.2130 |
| Output-Gap | 0.2065 | 0.0866 | 2.39 | 0.017*** | 0.0368 | 0.3762 |
| Ln_Revenue | 0.6749 | 0.0395 | 17.09 | 0.000*** | 0.5975 | 0.7522 |
| Ln_Beta | -0.6241 | 0.4285 | -1.46 | 0.1450 | -1.4639 | 0.2157 |
| Constant | 6.6992 | 0.7405 | 9.05 | 0.000*** | 5.2479 | 8.1505 |
| Dependent: Cash-on-Market | | | | | | |
| Ln_Valuation | 25.9870 | 18.9107 | 1.37 | 0.1690 | -11.0772 | 63.0513 |
| Country-Risk-Premium | -10355.8800 | 6041.6010 | -1.71 | 0.0870* | -22197.2000 | 1485.4420 |
| Output-Gap | 108.1019 | 29.6101 | 3.65 | 0.000*** | 50.0672 | 166.1366 |
| Constant | 981.7159 | 332.3529 | 2.95 | 0.003*** | 330.3162 | 1633.1160 |
| Dependent: Country-Risk-Premium | | | | | | |
| Output-Gap | -0.0024 | 0.0002 | -13.59 | 0.000*** | -0.0028 | -0.0021 |
| Constant | 0.0062 | 0.0003 | 22.05 | 0.000*** | 0.0057 | 0.0068 |
| | | chi2(7) | 14.24 | | Prob. > chi2 | 0.0471 |

| Wald tests for equations | | | |
|---|---|---|---|
| observed | chi2 | df | p |
| Ln_Valuation | 425.89 | 5 | 0.000*** |
| Cash-on-Market | 44.67 | 3 | 0.000*** |
| Country-Risk-Premium | 184.72 | 1 | 0.000*** |

| Equation-level goodness-of-fit | | | | | | |
|---|---|---|---|---|---|---|
| Dependent Variable | Fitted Variance | Predicted Variance | Residual Variance | R-Squared | mc | mc2 |
| Ln_Valuation | 9.400332 | 4.230804 | 5.029342 | 0.4649825 | 0.6819868 | 0.465106 |
| Cash-on-Market | 659319.4 | 54261.48 | 588650.1 | 0.1071852 | 0.3302521 | 0.1090664 |
| Credit-Risk-Premium | 0.0000438 | 0.0000116 | 0.0000322 | 0.2643713 | 0.5141705 | 0.2643713 |
| Overall | | | | 0.5910921 | | |

Figure 3 displays the SEM path-diagram summarizing the regression results of Tables 10 and 11 in Panels A and B respectively. In comparison, while both structural-equation model regressions demonstrate six statistically-significant coefficients, Panel A is both the more parsimonious model, with six of eight coefficients achieving significance, and also the model with lower overall P-values for its significant coefficients. Meanwhile, Panel B's circular component does not achieve statistical significance. Altogether, this corroborates the goodness-of-fit values in Tables 10 and 11, indicating that the non-circular structural-equation model regression has the stronger explanatory-power.

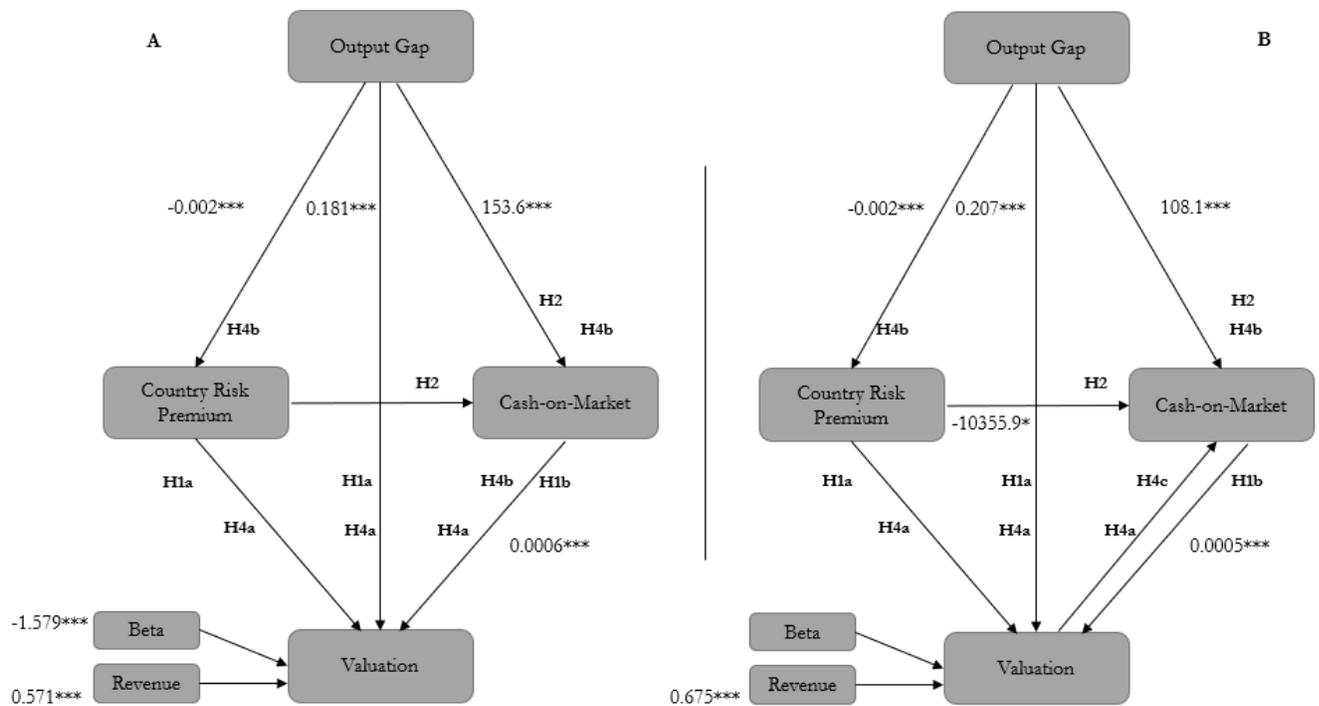

*Figure 3: SEM Path-Diagrams for Non-Circular and Partially-Circular SEM Regressions*

In principle, taking both the Sobel test and regression findings into account, we confirm the existence of a partial-mediation relationship between output-gap and startup-valuation. This means that business-cycles impact startup-valuation by impacting not only startups directly (i.e., in terms of revenues, business-relationships, consumer demand, and asset-values), but also indirectly via cash-on-market and country-risk-premium impacts.

Concerning endogeneity, our findings indicate that while investments significantly-drive valuations, valuations do not significantly drive valuations within our dataset, although the valuation-cash-on-market-relationship is nearly-significant, with a p-value of 0.169. Stated otherwise, the findings uphold the model described in Panel A rather than that described in Panel B, meaning that endogeneity is driven by variable-structure and hidden-variable-bias, rather than by direction-of-causality.

Compared to saturated direct-regression models (Gefen et al., 2011), using OLS and fixed-effects approaches, modeling the relationship between startup-valuation and business-cycle as a partial-mediation relationship nearly doubles model goodness-of-fit, while partial-mediation structure as a whole is both separately and jointly-significant. This approach not only successfully maps direct and indirect causality-structures linking business-cycles, VC-markets, and startup-valuations, but also a substantial goodness-of-fit improvement.

## 6. Findings and Implications

Market-conditions demonstrably influence startup-valuations. Primarily, we reach two major theoretical findings. First, our finding that relationships between startup-valuations and macroeconomic market-conditions can be modelled as a complex, multistep, and partially-indirect relationship echoes similar relationships found in VC syndication markets (Verwaal et al., 2010), international trade relationships (Troncoso and Gomes, 2020), and SME credit markets (Palazuelos et al., 2018). While such relationships are traditionally discovered in marketing and psychology, or in macroeconomic settings, it is likely going forward, that entrepreneurial finance will also find similar relationships with increasing frequency. Second, we find that impact on startup-valuations is primarily driven by business-cycle, while other macroeconomic market conditions such as country-risk-premium and cash-on-market, act as mutually-independent business-cycle distribution-channels rather than valuation-drivers in their own right. Therefore, connections between micro-level startup-markets and entrepreneurial finance with macrofinancial and macroeconomic market-conditions and market-theory, a clear theoretical development need identified by Budhwar et al. (2022), can be mapped and described as concretely-interrelated, while market-condition valuation-drivers themselves are more accurately described as valuation-channels rather than valuation-factors.

**Functional Form**

Our findings demonstrate that business-cycle not only drives valuations via direct-impact, but also drive macroeconomic and macrofinancial factors, further impacting startup-valuations indirectly. Mechanically speaking, we find that output-gap influences startup-valuations via both direct and indirect-effects. Specifically, this occurs via output-gap's impact on credit-risk-premium and domestic cash-on-market. While multicollinearity testing demonstrates that each of these valuation-impacts behave as individual and separate valuation-channels, with independent magnitudes of effect, they all establish that the valuation-process is established via intermediate-variables and valuation-channels rather than solely by valuation-factors in the classical sense.

Our analysis-results obtained through structural-equation modeling contribute to entrepreneurship-literature by describing an empirically-supported model demonstrating how business-cycles impact startup-valuations both directly and via multiple independent macro-level valuation channels.

Comparing direct-valuation effects with indirect-effect-models, we find that partial-mediation indirect-effect approaches with have substantially stronger goodness-of-fit and more accurately explain startup-valuation-impacts of market-conditions. Concerning endogeneity, while no convincing empirical-evidence of circular-relationship emerges, considering cash-on-market impacts on the SEM model outlined in Table 10, as well as the long-delay impacts demonstrated in the impulse-response-function and Table 9's optimal-lag tests, we find evidence for two-stage causality-structure, which fits the Wooldridge (2010) description of endogeneity driven by hidden-variable-bias.

Fundamentally, this means that while concrete valuation-relationships between startup markets and the macroeconomic-level exist, the valuation-relationship is demonstrably complex and indirect, featuring both direct and indirect valuation-channels. While this implies that startup-valuations are concretely driven by cyclical and macroeconomic factors, it also implies that the way startup-valuations might respond to changes in the business-cycle or macrofinancial-landscape may be complex, interactive, and multistage in nature, as effects are transmitted via the independent valuation-channels and are influenced by channel-specific mediating-variables.

**Common Threads and Key Variables**

The most consistently-powerful explanatory-factors are cyclical-indicators, country-risk-premiums, and cash-on-market. The reason for consistency of this explanatory-power is the partial-mediation structural-relationship between business-cycle indicators, financial-market-conditions, and startup-valuations.

In addition, examining interaction-effects indicates that during booms, valuation-discounts due to financial-market-conditions sectoral-beta and country-risk-premium are multiplied, further exacerbating the pro-cyclicality of startup-valuations. Valuations are pro-cyclical.

Business-cycle indicators are the most influential valuation-determinants modelled in our study due to their combination direct and indirect deterministic effect. Output-gap, a key business-cycle indicator, has on startup-valuations, which is expressed both directly, (ie, influencing revenues, asset-valuations, business relationships, and market-demand), but also indirectly (via the effect business-cycles have on country-risk-premiums and domestic cash-on-market). Not only are startup-valuations directly driven by output-gaps, as investors and entrepreneurs both enter booming markets, but also domestic financing conditions (as modelled in Figure 1).

Meanwhile, cash-on-market is found to act as a moderating variable, directly-impacting startup-valuations, while itself being directly impacted by business-cycles. In terms of overall direct-effect, results indicate that cash-on-market is the strongest macro-level factor. While world-cash-on-market has substantial explanatory-power, indicating substantial presence of cross-border-investment within our dataset, this strong cross-border-investment effects may be specific to the European common-market. Domestic cash-on-market, which is nearly as strong an indicator as world-cash-on-market, is itself influenced by both output-gap and tax-rates, as demonstrated in Table 5. Moreover, we demonstrate both via Sobel testing and via the SEM regression results that cash-on-market behaves as business-cycle transmission-channel vis-à-vis startup valuations.

On the other hand, regression-results indicate that dry powder, while a statistically-significant factor in some tables, is of limited explanatory power, given its extremely small significant coefficients, indicating that even large differences in country-level dry powder account for relatively-small increases in valuation, on average. Additionally, its non-significant Sobel test indicates that unlike cash-on-market, dry powder is far-enough removed from the business-cycle that it does not act as a business-cycle transmission-channel.

Like domestic cash-on-market, country-risk-premiums also act as moderating-variables, directly-impacting startup-valuations, while also being directly impacted by business-cycles. While baseline-models outlined in Table 2 demonstrate that country-risk-premiums negatively-impact startup-valuation, as per DCF valuation-approaches, as well as negatively-impacting cash-on-market outlined in Table 5, Tables 6 and 7 demonstrate that fixed-effects eclipses or partially-eclipses statistical-significance of country-risk-premiums depending on the categorical-variable examined, while Tables 10 and 11 demonstrate that country-risk-premiums are influenced by business-cycle conditions, and therefore act as a mediating-variable in the relationship between business-cycle and startup-valuations. In principle, country-risk-premiums proxy and signal macro-level market-conditions, corroborating Fama (1970)'s efficient-market concept.

**Hypotheses**

Regression-results indicate that Hypotheses 1 and 2 are clearly-demonstrated, with valuations being decisively-influenced by output-gaps, as per H1a, as well as both domestic and global cash-on-market, as per H1b. That global cash-on-market demonstrates stronger explanatory-power is indicative of extensive cross-border VC-investment driving startup-valuations. Given that our dataset is drawn from the EU and EEA prior to Brexit, this result is intuitive. The nuances and contours of this demand further examination in future research.

Meanwhile, Hypothesis 3, is partially-established. While fixed-effects findings tell a slightly-different story, it is likely the case that our dataset's categorical-variables are effectively proxying or substituting the impact of financial conditions on the business environment. However, it does not appear to be the case that there exists pockets or subsets of the dataset, for which the business-cycle's valuation-impact diverges substantially from the general situation.

Hypothesis 4 in its non-circular form, is also demonstrated. That can be established given that startup-valuations are impacted by domestic cash-on-market, which, in turn is impacted by output-gap. This is corroborated by the findings of Tables 3, 4, 10 and 11, by both regression-results and Sobel-testing.

Startup-valuations depend on business-cycles not only because of their direct-effects on the business-environment, but also because of their indirect effects via their impact on the financial-landscape, which impacts the business-environment in turn. Our findings relating to Hypothesis 4 add a layer of complexity to the narrative explaining how and why business-cycles influence startup-valuations. Domestic cash-on-market levels are both pro-cyclical and highly-deterministic to startup-valuations.

Concretely, this means that because startup-valuations are driven by cyclical and macrofinancial factors via both direct and indirect-effects, and that the effect is generalizable across industries, markets, and deal-types, we have ground to challenge classical valuation-models as being oversimplified and perhaps unsuitable for startup-markets. Instead, a staged, segmented valuation-model, along the lines of the Berre-Le Pendeven Startup-Valuation Meta-Model might be more accurate for startup markets.

**Implications**

This study is the first to explore the indirect connection between startup premoney valuation, business-cycle, and intermediate macrofinancial variables in order to capture the indirect role of business-cycles on financial-intermediary valuation-channels. Our results indicate that venture capital markets are to be seen and complex and pro-cyclical nonbank financial intermediaries that loudly echo ongoing and developing macroeconomic trends, with architecture and mechanics of this echo occurring via complex and partially-indirect processes.

Building on arguments from studies describing the macrofinancial, macroeconomic, and business-cycle determinants startup and venture capital markets such as Bernoth and Colavecchio (2014) and Troncoso and Gomes (2020), as well as studies describing indirect deterministic-impacts in entrepreneurship and small-business studies including Verwaal et al. (2010) and Palazuelos et al. (2018)

we describe how business-cycles influence startup-markets via independent, multiple valuation-channels.

**Context Matters**

Beyond our key theoretical and empirical findings, our paper contributes to the entrepreneurial-finance field by asking and addressing complex questions of startup and VC-market data, which are macroeconomic and multidimensional in nature, and which may likely drive further interdisciplinary theoretical and empirical pursuits in this area.

Our findings demonstrate that context matters for the process of establishment of startup-valuations. Here, we add that the manner by which context plays its deterministic role is itself multi-layered and subject to external macroeconomic influence. Additionally, our rejection of circular-relationship-endogeneity as a source of contextual valuation-determinant influence establishes a clear direction-of-flow by which market-conditions impact startup-valuations. Furthermore, our paper contributes to the wider field of economics by asking whether and how financial intermediary markets connect the entrepreneurial landscape to the wider macroeconomy in ways which are more than just unidirectional. Herewith, we hold previous empirical studies demonstrating unidirectional of nonbank financial intermediary impact to have been something of an oversimplification.

## 7. Conclusions and Further Studies

In mapping the nuanced causal relationship connecting business-cycles VC markets and startup-valuations, this study makes a substantial contribution to the macrofinancial, entrepreneurial-finance and nonbank financial-intermediary fields. Furthermore, we respond to numerous identifiable gaps in the literature. These include the pronounced need for research tying together entrepreneurial-finance and macro-level economic theory identified by Budhwar et al. (2022) as well as the need for methodological-diversification in venture capital research identified by Berre and Le Pendeven (2022) and the need for insights on macro-level transmission roles of nonbank financial-intermediation as identified by European Central Bank (2021). Using an SEM approach, we address these gaps in the research by demonstrating how business-cycles, startup markets and venture-capital intermediary markets are related in nuanced detail.

Overall, while startup-valuations are demonstrably consistent with discounted-cashflow valuation approaches, both macro-level country-risk-premiums and venture capital markets influence startup-valuations. These in turn, are driven by business-cycles. Essentially, this means that startup-valuations are driven by reigning market-conditions as much as revenues and discount-rates. Our findings additionally demonstrate startup-valuations to be extremely pro-cyclical, because business-cycles influence startup-valuations both directly and indirectly. Nevertheless, while business-cycles influence startup-valuations both directly and indirectly, the evidence points to a non-circular relationship.

Crucially, SEM results indicate that whereas classically-established economic-theory (i.e., DCF, multiples) describes valuation as flowing from firm-level characteristics and financial market-conditions, startup-valuations may actually convey the multi-channel impact of macro-economic business-cycle conditions impacting the private-equity and VC markets directly and indirectly via multiple uncorrelated transmission-channels. The implication being that country-risk-premium, cash-on-market and possibly other market-conditions essentially proxy business-cycle-conditions.

Our findings have implications for both investors and economic policy makers. For the latter, the implication is that because entrepreneurial finance markets are strongly procyclical, it is possible to influence entrepreneurial valuations at the national level by means of policy measures which impact the business-cycle, such as monetary policy or macro-level fiscal policy. Furthermore, detailed mapping of the business-cycle's transmission channels allows the crafting of bespoke entrepreneurial finance policy. Meanwhile, for investors, the implication of this study's findings relating to both procyclicality and variable interrelatedness can be realized on both international venture finance markets by offshore VC investors examining a country's startup market, and by long-term VC investors who depend on timing the market. While the former could make use of our findings in order to optimize the choice of overseas target market, the latter could make use of them to optimize their market-timing, even regarding the timing-investments constraints of most of the VC firms.

**Research and Further Studies**

Results presented in this study point to several places where future VC and entrepreneurial-finance research might achieve substantial future contributions.

First, since macroeconomic and macrofinancial drivers have substantial valuation-impact, future studies examining cross-border VC deals by comparing valuation-impact by startup's home-jurisdiction impact to those investors' jurisdiction, as well as differences thereof, might uncover evidence of yield-chasing, stability-chasing, sectoral-cross-border effects, or even substantial analyzable fault-lines.

Second, this study's findings can be used to construct a VC market-model analogous to the quantity-theory of money, substituting money-supply's macro-level role with that of cash-on-market supply, supplementing empirical findings on the role of cash-on-market quantity with those of measures analogous to velocity, such as deals, exits, and investment durations.

Third, our fixed-effects findings indicate significant fault-lines characterized by data-subsets where startup-valuation-sensitivities to identified market-condition drivers diverge significantly. Because this indicates that categorical variables have substantial explanatory-power, research-possibilities exist, methodologically-speaking to explore categorical-variable valuation-impact using novel empirical approaches, such as non-parametric statistical techniques, qualitative-techniques, or hybrid-methodologies. Likewise, city-fixed-effects findings indicate that investigation into municipal-level market-conditions might yield detailed findings concerning valuation-impacts of local-level drivers.

# 9. Appendix I: Variable Definitions

Table 12 describes variables used, as well as variable-definitions and sources.

*Table 13: Variable Definitions*

| Variable | Description |
|---|---|
| Valuation | Pre-Money Startup-Valuation. EIKON, Early Metrics, Crunchbase |
| Revenue | Startup company-revenue: EIKON, Dun & Bradstreet, Zoominfo, |
| Beta | Unlevered sectoral-beta. NYU-Stern dataset |
| Country-Risk-Premium | Country-Level Country-risk-Premium. Moody's, NYU-Stern |
| Tax-Rate | Tax-revenue as % of GDP. World Bank Indicators |
| Output-gap | Deviations of actual GDP from potential GDP as % of potential GDP. OECD |
| Dry Powder | Country-level estimate. Amount of committed but unallocated VC capital on hand. Invest Europe |
| Cash-on-Market | Country-level total VC investments. OECD |
| World Cash-on-Market | Worldwide total VC investments. OECD |

# 10. Appendix II: Multicollinearity

## Are Output Gaps and Cash-on-Market Correlated? Causally Related?

Because multichannel cyclical valuation-impact is a key-contribution, multicollinearity-testing is necessary to establish statistical non-relatedness among the macro-level variables influencing startup-valuation. In Table 13, multicollinearity is investigated using variance inflation factor (VIF), conducted subsequent to a multivariate-regression using our primary macro-level variables, outlined in Table 3. This measures correlation and strength-of-correlation between explanatory-variables. Table 11 indicates non-correlation among macro-variables. No variable has a VIF-score surpassing 2.00, indicating mutual-independence of direct factors influencing startup-valuation, demonstrated in Figure 1 and Table 3.

*Table 14: Multicollinearity-Testing*

| Source | SS | df MS | Number of obs | 514 | |
|---|---|---|---|---|---|
| | | | F( 5, 508) | 90.25 | |
| Model | 2283.0547 | 5 456.610941 | Prob > F | 0 | |
| Residual | 2570.24153 | 508 5.05953057 | R-squared | 0.4704 | |
| Total | 4853.29624 | 513 9.46061644 | Adj R-squared | 0.4652 | |
| | | | Root MSE | 2.2493 | |
| Ln_Valuation | Coef. | Std. Err. t | P>t | [95% Conf. | Interval] |
| Ln_Revenue | 0.6714774 | .0395184 16.99 | 0.000 | 0.5938378 | 0.7491171 |
| Ln_Beta | -0.6011327 | .4294032 -1.40 | 0.162 | -1.444757 | 0.242492 |
| Country-Risk-Premium | -15.39393 | 17.77113 -0.87 | 0.387 | -50.30789 | 19.52003 |
| Output-Gap | 0.181132 | .0847502 2.14 | 0.033 | 0.0146279 | 0.3476361 |
| Cash-on-Market | 0.0007468 | .0001272 5.87 | 0.000 | 0.0004969 | 0.0009967 |
| Constant | 6.426215 | .7142596 9.00 | 0.000 | 5.022948 | 7.829481 |
| Variable | VIF | 1/VIF | | | |
| Output-Gap | 1.45 | 0.688246 | | | |
| Country-Risk-Premium | 1.4 | 0.712027 | | | |
| Ln_Revenue | 1.11 | 0.897474 | | | |
| Cash-on-Market | 1.08 | 0.921813 | | | |
| Ln_Beta | 1.07 | 0.933796 | | | |
| Mean VIF | 1.23 | | | | |

# 11. Appendix III: Growth-Cycle Coincident Indicator

## An Alternate business-cycle indicator

This appendix examines business-cycle impact using an alternate business-cycle indicator. Eurostat growth cycle coincident indicator (GCCI) indicates probability of slowdown in the macroeconomy, signals growth-cycle peaks and troughs, and is published for the Eurozone as a whole. What we observe is that OLS regressions demonstrate that valuation is directly impacted by the GCCI indicator, as is cash-on-market, but that SEM models indicate the impact significant on country-risk-premium, indicating that GCCI measures would describe business-cycle more as an indirect impact than a direct one.

*Table 15: Direct and Indirect Impact of Macro-level Startup Valuation-Drivers using GCCI*

**GCCI Business-Cycle Regressions**

| VARIABLES | (1) Ln_Valuation | (2) Ln_Valuation | (3) Ln_Valuation | (4) Ln_Valuation | (5) Ln_Valuation | (6) Ln_Valuation |
|---|---|---|---|---|---|---|
| Ln_Revenue | 0.6514*** | 0.6540*** | 0.6356*** | 0.7060*** | 0.6862*** | 0.6071*** |
|  | [0.034] | [0.034] | [0.035] | [0.037] | [0.036] | [0.034] |
| Ln_Beta | -1.0886*** | -1.3509*** | -1.1604*** | -0.8804** | -0.9448** | -0.9861*** |
|  | [0.388] | [0.383] | [0.387] | [0.406] | [0.390] | [0.374] |
| Country-Risk-Premium | -38.3685*** | -15.3681 | -38.587*** | -43.2772*** | -23.8422* | -41.5507*** |
|  | [13.092] | [13.622] | [15.048] | [13.048] | [14.386] | [13.232] |
| Tax Rate |  | 0.1130*** |  |  |  |  |
|  |  | [0.020] |  |  |  |  |
| GCCI |  |  | 0.49393** |  |  |  |
|  |  |  | [0.213] |  |  |  |
| Dry-Powder |  |  |  | 0.00000009** |  |  |
|  |  |  |  | [0.000] |  |  |
| Cash-on-Market |  |  |  |  | 0.0006*** |  |
|  |  |  |  |  | [0.000] |  |
| World-Cash-on-Market |  |  |  |  |  | 0.00003*** |
|  |  |  |  |  |  | [0.000] |
| Constant | 7.9171*** | 3.8202*** | 8.0147*** | 6.6447*** | 6.4131*** | 5.3569*** |
|  | [0.615] | [0.933] | [0.615] | [0.687] | [0.648] | [0.639] |
| Observations | 646 | 636 | 636 | 557 | 578 | 550 |
| R-squared | 0.41 | 0.44 | 0.42 | 0.45 | 0.48 | 0.53 |
| Adjusted R-squared | 0.408 | 0.438 | 0.412 | 0.447 | 0.473 | 0.526 |

Standard errors in brackets
*** $p<0.01$, ** $p<0.05$, * $p<0.1$

**GCCI Impact on Cash-on-Market**

| VARIABLES | (1) Cash-on-Market | (2) Cash-on-Market | (3) Cash-on-Market | (4) Cash-on-Market | (5) Cash-on-Market | (6) Cash-on-Market |
|---|---|---|---|---|---|---|
| GCCI | 920.1494*** | 912.9339*** | -20.1683 | 166.6458*** | 79.9328 | 3.5830 |
|  | [78.825] | [74.102] | [107.487] | [57.253] | [103.135] | [100.413] |
| Tax-Rate | -14.4339** | -28.7325*** | -14.3335** |  |  | -27.6117*** |
|  | [6.345] | [6.176] | [5.752] |  |  | [5.563] |
| Country-Risk-Premium |  | -40,519.8447*** |  |  | -29,451.9786*** | -36,781.4116*** |
|  |  | [4,545.210] |  |  | [3,980.538] | [4,010.077] |
| Cash on World Market |  |  | 0.0176*** | 0.01398*** | 0.0160*** | 0.0168*** |
|  |  |  | [0.002] | [0.0009] | [0.002] | [0.002] |
| Constant | 1,583.0907*** | 2,444.4569*** | -0.6567 | .0139845*** | -135.3966 | 862.4071*** |
|  | [231.023] | [237.692] | [244.460] | [83.264] | [160.296] | [246.928] |
| Observations | 599 | 599 | 571 | 786 | 576 | 571 |
| R-squared | 0.19 | 0.29 | 0.27 | .28 | 0.33 | 0.36 |
| Adjusted R-squared | 0.192 | 0.286 | 0.264 | .279 | 0.322 | 0.358 |

Standard errors in brackets
*** $p<0.01$, ** $p<0.05$, * $p<0.1$

| Structural equation model | Number of obs: | 1089 | | | | |
|---|---|---|---|---|---|---|
| Estimation method = ml | | | | | | |
| Log likelihood = -7103.2086 | | | | | | |
|  | Coef. | SE | z | P-Value | [95% Conf. | Interval] |
| **Dependent: Ln_Valuation** | | | | | | |
| Cash-on-Market | 0.0006 | 0.0001 | 5.7 | 0.0000 | 0.0004 | 0.0008 |
| Country-Risk-Premium | -26.7696 | 10.3907 | -2.58 | 0.0100 | -47.1349 | -6.4042 |
| GCCI | -0.0810 | 0.1817 | -0.45 | 0.6560 | -0.4371 | 0.2752 |
| Ln_Revenue | 0.5917 | 0.0301 | 19.65 | 0.0000 | 0.5327 | 0.6507 |
| Ln_Beta | -1.3501 | 0.3361 | -4.02 | 0.0000 | -2.0089 | -0.6913 |
| Constant | 7.9103 | 0.5429 | 14.57 | 0.0000 | 6.8464 | 8.9743 |
| **Dependent: Cash-on-Market** | | | | | | |
| Country-Risk-Premium | -22955.9900 | 3826.5600 | -6 | 0.0000 | -30455.9100 | -15456.0700 |
| GCCI | 550.2820 | - | - | - | - | - |
| Constant | 1244.5150 | 41.5438 | 29.96 | 0.0000 | 1163.0910 | 1325.9400 |
| **Dependent: Country-Risk-Premium** | | | | | | |
| GCCI | -0.0015 | 0.0005 | -2.99 | 0.0030 | -0.0025 | -0.0005 |
| Constant | 0.0093 | 0.0003 | 33.38 | 0.0000 | 0.0087 | 0.0098 |
|  | | chi2(10) | 21.62 | | Prob. > chi2 | 0.0171 |